\documentclass[preprint,11pt]{elsarticle}
\usepackage[letterpaper,margin=0.9in]{geometry}

\biboptions{numbers,sort&compress}
\usepackage{textcomp}
\usepackage[T1]{fontenc}
\usepackage{amssymb}
\usepackage{gensymb}
\usepackage[utf8]{inputenc}
\usepackage{newunicodechar}
\newunicodechar{−}{-}
\usepackage{amsmath,amssymb}
\usepackage{bm}
\usepackage{changepage}
\usepackage{textcomp,marvosym}
\usepackage{xcolor}
\usepackage{microtype} 
\usepackage{xcolor}
\usepackage{nameref,hyperref}
\usepackage[right]{lineno}
\usepackage[table]{xcolor}
\usepackage{array}
\usepackage{textcomp}
\usepackage{amssymb}
\usepackage{subcaption}
\usepackage{lipsum}
\usepackage{multicol}
\usepackage{graphicx}
\usepackage{amsmath}
\usepackage[linesnumbered,ruled,vlined]{algorithm2e}
\usepackage{lineno}
\usepackage{comment}
\usepackage{booktabs}
\usepackage{subcaption}
\usepackage{graphicx}
\usepackage{caption}
\usepackage{array}
\usepackage{placeins}
\usepackage[utf8]{inputenc}
\usepackage{lineno}
\begin{document}
\begin{frontmatter}

\title{A two-parameter, minimal-data model to predict dengue cases: the 2022–2023 outbreak in Florida, USA}

\author[label1]{Saman Hosseini\corref{corrr1}}

\author[label2]{Lee W. Cohnstaedt}
\author[label1]{Caterina Scoglio}

\affiliation[label1]{organization={Department of Electrical and Computer Engineering, Kansas State University},           
            city={Manhattan},
            state={KS},
            country={USA}}

\affiliation[label2]{organization={Foreign Arthropod-Borne Animal Diseases Research Unit, National Bio- and Agro-defense Facility, USDA ARS},
            city={Manhattan},
            state={KS},
            country={USA}}

\cortext[corrr1]{Corresponding author \\ Email address: shosseini@ksu.edu}

\begin{abstract}
Reliable and timely dengue predictions can provide an actionable lead time for targeted vector control and clinical preparedness, reducing preventable disease and health-system costs in at-risk communities. 
However, many forecasting approaches depend on site-specific covariates and entomological surveillance, which limits portability to data-sparse settings.
In this work, we mathematically prove that a parabolic ICC structure, previously established for the basic SIR model, also holds in a substantially more complex model: a two-population (human–mosquito) four-serotype dengue transmission model with primary and secondary infections and mild/severe disease classes.
To predict the number of new cases, we propose a data-parsimonious (DP) framework built on the incidence–cumulative cases (ICC) curve that requires only the human incidence time series of the target season and estimates only two parameters, thereby reducing estimation noise and computational burden.
We further develop a Bayesian extension that accounts for case-reporting and fitting uncertainty, producing calibrated predictive intervals. 
The Bayesian model yields improved predictive performance compared to the parabolic ICC model.
We evaluate the framework for dengue outbreaks in Florida in 2022–2023, where standardized clinical tests and reporting support accurate case determination.

\end{abstract}



\end{frontmatter}




\section{Introduction}
\label{sec1}
Dengue is the most prevalent arthropod-borne viral infection in the human population \cite{simmons2012dengue,ross2010dengue}, transmitted by mosquitoes of the genus \textit{Aedes} \cite{schaffner2014dengue}.
It comprises four antigenically distinct serotypes (DENV$-$1, DENV$-$2, DENV$-$3 and DENV$-$4).
It places roughly half of the world's population at risk and causes an estimated 100$-$400 million infections each year \cite{WHO_dengue_factsheet}.
Transmission occurs in tropical and subtropical regions, particularly in urban and peri-urban settings.
Most infections are asymptomatic or mild, but a proportion progresses to severe disease that can be fatal \cite{paixao2015trends}.

Although dengue vaccines are available, there is no proven, specific antiviral therapy for dengue \cite{lai2017pharmacological}.
In addition, the performance of the vaccine has important limitations: its effectiveness varies by serotype, usually stronger for DENV$-$ 1 $/$ 2 and weaker or uncertain for DENV$-$ 3 $/$ 4, and its protection is generally higher in previously exposed individuals (seropositive) than in seronegative recipients \cite{thomas2023new}.
Given the limitations of dengue vaccination, population-level measures, such as vector control programs and public health alerts, remain essential to reduce transmission \cite{knerer2015impact,rather2017prevention}.
Consequently, accurate forecasting is integral to effective mitigation.
Accurate forecasts enable vector control agencies to reduce transmission risk by reducing mosquito populations through larviciding and adulticiding, deploying biotechnological approaches (e.g., sterile insect technique) \cite{ranathunge2022development}, or combining these strategies, while timely public health advisories increase the community's uptake of personal protective measures.
Thus, reliable forecasting moves mitigation from reaction to prevention, enhancing cost efficiency and health outcomes.

To operationalize such forecasting, models rely on measurable drivers drawn from multiple data streams.
There are various data sources to predict the risk of dengue transmission to humans.
A strong correlation between dengue incidence and \textit{Aedes} spp. abundance has been shown \cite{bowman2014assessing,sasmita2021ovitrap,ong2021adult}; thus, a large number of models incorporate entomological drivers to predict the risk of transmission from the mosquito to the human population.
Consequently, due to the correlation between mosquito abundance and climatological variables such as temperature, rainfall, and humidity \cite{lega2017aedes,tran2013rainfall,serpa2013study,wang2023spatial} climatological drivers are also widely used in dengue forecasting models.

In general, the data used for the models can be classified as direct and indirect factors \cite{siriyasatien2018dengue}.
Direct factors are those that immediately act on vector biology and dengue transmission by altering the mosquito life cycle or virus dynamics; they include climate and weather drivers (rainfall, temperature, humidity, drought, El Niño) \cite{buczak2012data,naish2014climate,gharbi2011time,mclennan2014complex,karim2012climatic,wu2009higher,reiter2003texas,wongkoon2013weather,xu2014statistical,machado2012empirical,runge2008does}, mosquito density measures (positive containers in houses inspected; larval, pupal, and adult counts) \cite{focks1997pupal,seng2009pupal,pham2011ecological,aburas2007aburas,strickman2003dengue,thongrungkiat2012natural,ponlawat2007age}, dengue virus serotypes, vertical transmission \cite{karim2012climatic,limkittikul2014epidemiological}, and human biting rate \cite{chompoosri2012seasonal,thavara2015biology}.
Indirect factors are contextual conditions that shape outbreaks without directly changing larval counts: geography and topography \cite{cdc2007dengue,knowlton2009fever} and urban–rural setting \cite{rund2019rescuing,biswas2014dengue}; spatial and spatiotemporal patterns \cite{yu2014online,c2015surveillance,thiruchelvam2018correlation}; population movement and trade \cite{beatty2005travel,baaten2011travel,ratnam2013dengue,hanna2006multiple,teichmann2004dengue,reiner2014socially,gardner2012predictive}; environmental, land-use and sanitation conditions; and host immunology and vaccination landscape.

The combination of direct and indirect factors has been used to establish predictive frameworks to forecast risk at different locations.
Hii et al. (2012) built a weather-based dengue forecasting model for Singapore using Poisson regression with temperature and rainfall modeled as piecewise linear (spline) effects over long lag windows, plus AR lags, trend, seasonality, and a population offset \cite{hii2012forecast}.
Descloux et al. (2012) developed a support vector machine (SVM) model for dengue outbreak forecasts for Noumea (New Caledonia) using epidemiological surveillance data, local meteorological variables, and large-scale climate indices \cite{descloux2012climate}.
Lowe et al. (2014) used a Bayesian negative binomial GLMM for monthly dengue forecast in Brazil with fixed effects for 3-month temperature and precipitation anomalies, altitude, population density, and 4-month lagged dengue risk, plus seasonal random effects and zones varying \cite{lowe2014dengue}. 
Shi et al. (2016) trained 12 horizon-specific LASSO models \citep{ranstam2018lasso} to predict Singapore’s weekly dengue 1–12 weeks ahead using recent case counts, vector breeding indices, meteorological variables and population statistics \citep{shi2016three}.
Lowe et al. (2016) developed a Bayesian negative binomial spatiotemporal GLMM for Brazil, driven by seasonal climate forecasts (3-month temperature/precipitation anomalies), population density, altitude and a 4 month lagged dengue signal, with spatial effects and zone-varying seasonality; generated 3–4 month probabilistic risk maps of 3-4 months that outperformed a seasonal baseline \citep{lowe2016evaluating}.
Alkhateeb et al. (2018) described a probabilistic forecasting system, deployed in Brazil, Malaysia, and Mexico, that uses logistic regression in weekly surveillance, meteorological and entomological indicators to trigger alerts once the predicted outbreak probability exceeds a predefined threshold \cite{hussain2018early}.
Additional models from various locations have used different drivers and methods for risk prediction \cite{aburas2010dengue,adde2016predicting,gultekin2010neural,edussuriya2021accurate,buczak2018ensemble,siriyasatien2016analysis,salim2021prediction,deb2017ensemble,majeed2023prediction,ramadona2016prediction,althouse2011prediction,bal2020modeling,chen2018neighbourhood}.

These approaches have advanced operational forecasting in diverse settings; nevertheless, several limitations remain.
Although many models perform well on timing, trajectory, and severity, they are often location-specific and fail to generalize to other settings. In particular, reliance on direct factors, such as entomological surveillance restricts use to sites where such data exist. Moreover, predictors that are strong at one site may be weak elsewhere; for example, humidity may be a primary driver in one location but show no correlation with outbreaks in another.
In addition, in many settings, there are no accurate statistics on human cases, which means that case reporting is subject to substantial uncertainty for a variety of reasons.
Beyond limited generalizability, two additional issues arise: parameter estimation adds noise and model selection is time-consuming and computationally intensive. Finally, many models are inapplicable for forecasting in locations without an outbreak history (newly infected locations). 

To address these limitations, we developed and evaluated a DP framework based on the incidence versus cumulative cases (ICC) curve method.
In this work, we provided mathematical proof that the parabolic ICC structure, previously established for the basic SIR model \cite{lega2016data,lega2021parameter}, remained valid in a substantially more complex model: a two-population (human–mosquito), four-serotype dengue transmission model with primary and secondary infections and classes of mild/severe disease.

The only two parameters of this model are estimated from human incidence data, reducing the noise of estimation.
Moreover, because the only required input is the time series of new human cases in the target year, the approach is usable in data-limited settings without entomological surveillance and with minimal historical information. 

Finally, based on the theory of the ICC framework, we constructed a Bayesian predictive model that incorporates uncertainty in reported cases and in parameter estimation to enhance reliability and deliver narrower uncertainty intervals.
We demonstrate the performance of the proposed framework by forecasting dengue risk in Florida during 2022–2023.
Our results showed that the Bayesian model achieved better accuracy compared to the parabolic ICC model.

In Section 2, we describe the data, study location, present the dengue compartmental model used to mathematically prove that the parabolic ICC relation holds for dengue, and outline the predictive framework, including point and Bayesian prediction and the metrics used to evaluate accuracy.
Section 3 presents results separately for each outbreak (point and Bayesian predictions).
Section 4 provides a discussion of the results of the model, including the interpretation of the result in light of the ecology of dengue in Florida, operational and public-health implications, and limitations with directions for future work.

In addition, this paper includes the supplementary material provided in two parts. The theoretical material is included as an Appendix at the end of the manuscript (in the same PDF as the main paper), while Supplementary.docx contains additional plots, tables, and related supporting material, in order to reduce the length of the main manuscript.

\section{Material and methods}
\noindent In this section, we describe the data sources, study location, disease model, predictive models, and the methods and metrics used to evaluate the proposed framework.
\subsection{Data}\label{sec: data}
For this study, we used data on locally acquired dengue cases in Florida, as reported by the US Centers for Disease Control and Prevention (CDC) for the years 2022 and 2023. The CDC reported locally acquired and travel-associated (non-local) cases separately; accordingly, our analysis uses only locally acquired case counts.
The two years were combined into a continuous data set that spanned a total of 104 weeks (Figure \ref{fig:Florida 2022-23-seasons}).
In Florida, suspected dengue is clinically evaluated and confirmed by laboratory tests.
For acute illness (0-7 days after the onset of the symptom), clinicians are advised to order both a nucleic acid amplification test (RT-PCR/NAAT) and an IgM antibody test because the combination detects more cases than either test alone.
Specimens are submitted, after coordination with the county health department, to the Florida Department of Health Bureau of Public Health Laboratories (BPHL) or a CLIA-certified commercial laboratory, following Florida Department of Health procedures \cite{FDA_CLIA_2023}. 
Acute serum is shipped refrigerated (2 to 8 ° C) or frozen ( -20 ° C), with acute samples often shipped frozen on dry ice.
These standardized testing and reporting requirements strengthen the determination of cases in Florida; consequently, the observed dengue case counts used in this study are likely to be accurate.
At the same time, the underlying true incidence is not directly observable, since asymptomatic and otherwise unreported infections are not captured  \cite{FDOH_BPHL_DENV_PCR_2023,CDC_Molecular_Tests_Dengue_2025,CDC_Clinical_Testing_Dengue_2025} .
The raw data, number of new cases for each week, can be found in the Supplementary X.

\subsection{Study location}
Florida serves as a practical case study for evaluating our model in a real-world setting of locally acquired dengue. 
The state spans subtropical to tropical climates on a peninsula bordered by the Gulf of Mexico, the Atlantic Ocean, and the Straits of Florida, with southern counties supporting conditions conducive to year-round arboviral transmission \cite{wilke2019community}. 
Although most U.S. dengue cases are travel-associated, sustained local transmission has been documented in Florida, particularly in the south \cite{stephenson2021geographic}, and locally acquired infections have increased in recent years, with more than 60 locally acquired cases reported in 2022 and nearly 200 in 2023 in several counties. 
This pattern of recurrent and expanding local transmission raises concern that parts of Florida may be on a path toward endemic dengue transmission if current trends continue.
At the same time, Florida can still be regarded as a relatively recently affected region, with limited long-term historical data on autochthonous dengue cases. From a modeling standpoint, this combination of emerging endemicity, constrained historical information, and robust surveillance makes Florida a particularly proper setting in which to evaluate and stress-test our DP model, which relies solely on reported human cases.

\subsection{Disease model} \label{sec: compartmental model}
Various compartmental differential equation models with varying levels of detail have been used to capture dengue transmission mechanisms.
In this study, we adopt a two-population (human–mosquito), four-serotype dengue transmission model with primary and secondary infections and mild/severe disease classes to characterize transmission and mathematically show that the parabolic ICC relation holds for dengue.
This model is a more complete four-serotype extension of the two-serotype framework introduced by Nuraini et al. (2007) \cite{Nuraini2007}.
We emphasize that this compartmental model is not used for prediction in this work; it is used solely to provide a mathematical justification for the parabolic incidence–cumulative cases relationship in dengue.
The differential equations that govern the model are presented below.
The compartments in this model represent different stages of infection status in both human and mosquito populations.
Taking a dot to denote the time derivative ($\frac{dy}{dt}\Rightarrow \dot{y}$), the compartments are defined as follows:

\noindent
\begin{minipage}[t]{0.49\textwidth}
\[
\begin{aligned}
\dot S \;&=\; -S\sum_{i=1}^{4}\lambda_{H,i},\\
\dot I_i \;&=\; S\lambda_{H,i}-\gamma I_i,\\
\dot R_i \;&=\; \gamma I_i - R_i\sum_{\substack{j=1\\ j\neq i}}^{4}\sigma_j\lambda_{H,j},\\
\dot Y_i \;&=\; (1-q)\sigma_i\lambda_{H,i}\sum_{\substack{j=1\\ j\neq i}}^{4}R_j-\gamma Y_i,
\end{aligned}
\]
\end{minipage}\hfill
\begin{minipage}[t]{0.49\textwidth}
\[
\begin{aligned}
\dot D \;&=\; q\sum_{i=1}^{4}\sigma_i\lambda_{H,i}\sum_{\substack{j=1\\ j\neq i}}^{4}R_j-\gamma D,\\
\dot R \;&=\; \gamma\Big(\sum_{i=1}^{4}Y_i + D\Big),\\
\dot S_M \;&=\; -S_M\sum_{i=1}^{4}\lambda_{M,i},\\
\dot V_i \;&=\; S_M\lambda_{M,i}, \qquad i=1,2,3,4.
\end{aligned}
\]
\end{minipage}

\noindent The model has been described in detail in Appendix A.

\subsection{Predictive model}
In this section, we describe the ICC relation and explain how ICC curves can be used to predict the epidemic trajectory.
We also define the censored Poisson model based on the parabolic ICC relation to predict the number of weekly cumulative cases.
Finally, we describe how to establish the Bayesian model grounded in the theory of the parabolic ICC reflation that counts for error in data reporting and curve fitting process.

\subsubsection{Predictive ICC-curve method for dengue}\label{forecast model}
We first introduce the following lemma. This result provides the key analytical tool for predicting the number of new cases, i.e., the epidemic trajectory.

\vspace{2mm}
\noindent \textbf{Lemma}. For the two-population, four-serotype dengue compartmental model described in Section~\ref{sec: compartmental model}, the incidence can be expressed as an inverse parabolic function of the cumulative case count.
\begin{flalign}\label{logistic_eq}
&\dot{C} \propto C\,(L_0 - C) 
\Rightarrow \dot{C}=W\,C\,(L_0-C)
  =W\,L_0\,C\left(1-\frac{C}{L_0}\right),
\end{flalign}

\noindent where \(L_0\) denotes the final epidemic size and \(W\) represents the intrinsic growth–rate constant.

\noindent The detailed proof is given in Appendix B.

\noindent The general form of Equation~\ref{logistic_eq} highlights two key points: first, it represents a differential equation that describes logistic growth over time; and second, the rate of change (incidence) is a quadratic function, in the form of an inverted parabola, with respect to the cumulative number of cases.
These two important features will provide a method for predicting the number of new cases during a given dengue outbreak.

\begin{figure}
    \centering
    \includegraphics[width=0.5\linewidth]{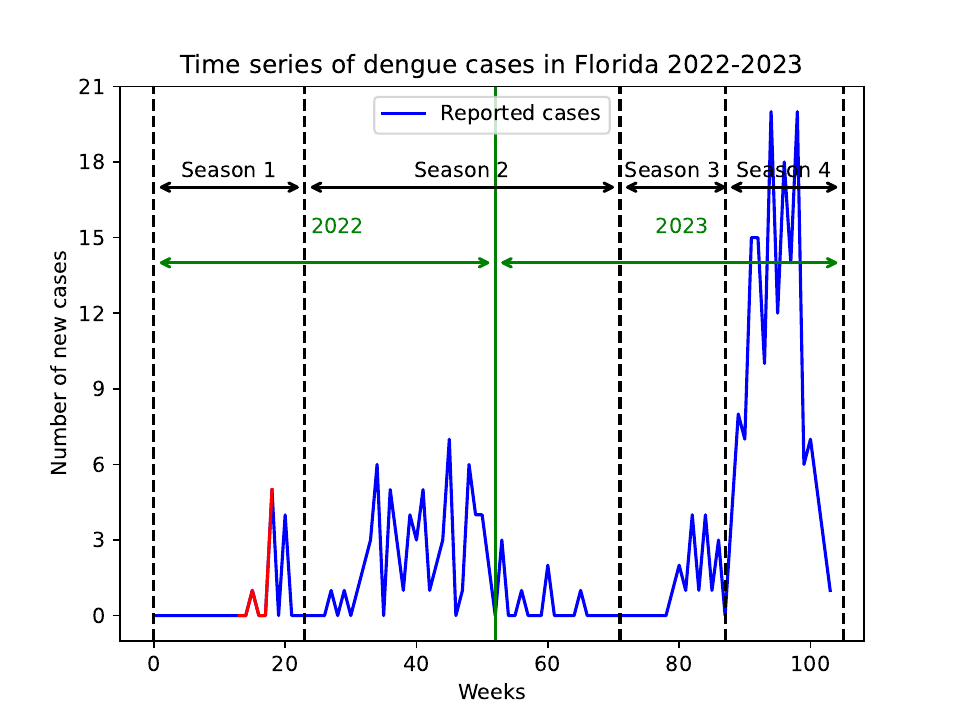}
    \caption{Time series of new cases in Florida for 2022-2023, highlighting the outbreak seasons. }
    \label{fig:Florida 2022-23-seasons}
\end{figure}

\noindent From a statistical perspective, the result means that the epidemic trajectory can be defined by a logistic growth curve. 
\begin{equation}\label{cumulative_cases}
C_t = \frac{L_0}{1 + \exp\{-\delta(t - \mu)\}},
\end{equation}
where \(C_t\) denotes the cumulative number of cases reported at time \(t\), \(L_0\) is the final size of the outbreak, \(\mu\) is the mean distribution and \(\delta\) controls the growth rate.
The cumulative curve (\ref{cumulative_cases}) can be written as the multiplication of two terms of $L_0$ and $\frac{1}{1 + \exp\{-\delta(t - \mu)\}}$. The second part is a cumulative distribution function (CDF) of a logistic random variable ($T$), so $C_t = L_0 P(T\leqslant t)$.
Importantly, for a given dengue outbreak, the time to infection \(T\), the time at which an individual of the final infected population becomes infected, is distributed as a logistic random variable.
This formulation offers a probabilistic interpretation of the epidemic curve, where \(\mu\) and \(\delta\) represent the location and scale parameters of the underlying logistic distribution.

\subsubsection{Outbreak prediction using ICC curve method} \label{point prediction}
\begin{figure}
  \centering
  \includegraphics[width=0.7\linewidth]{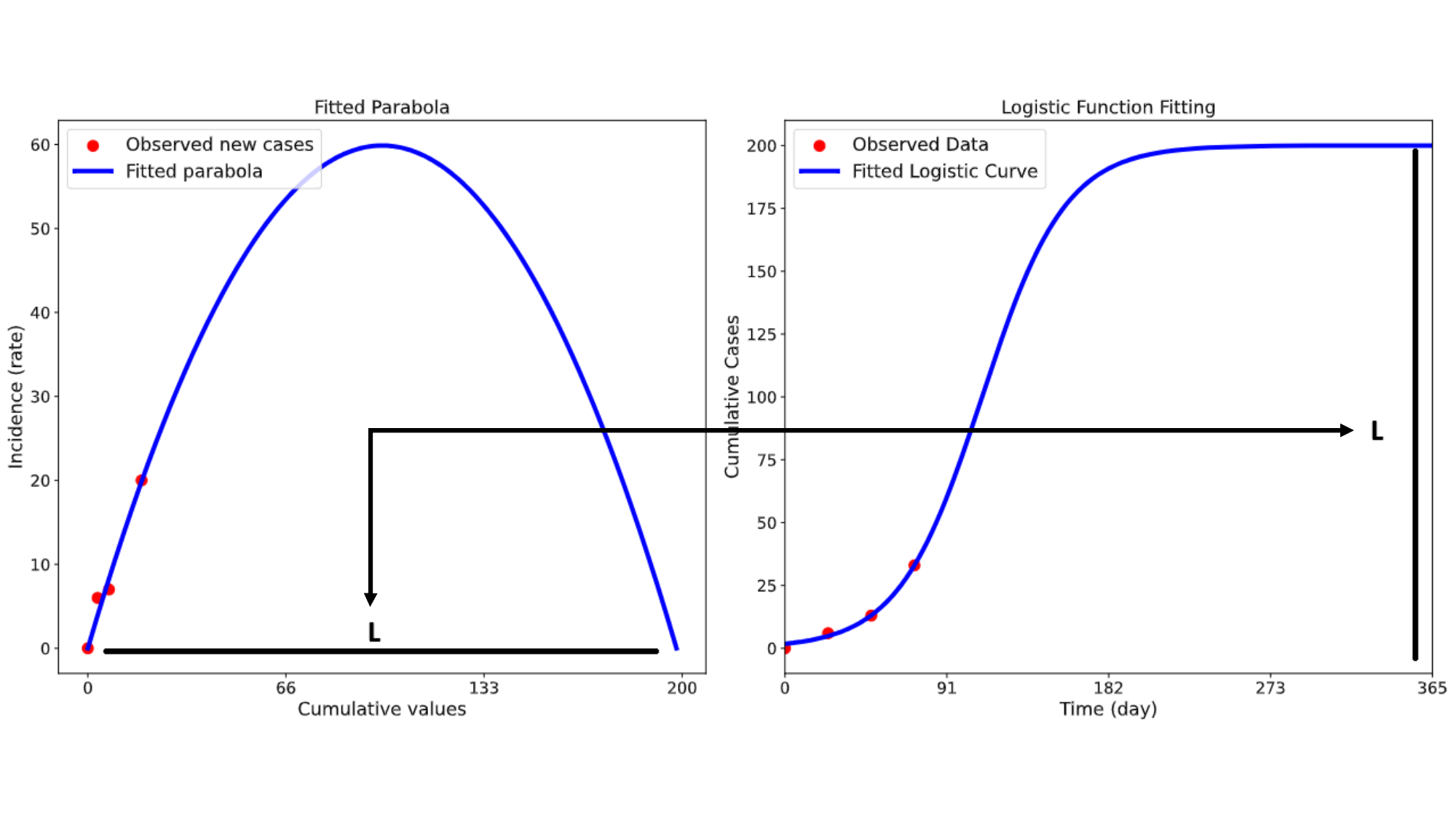} 
  \caption{Fitted parabola and logistic function to the data at the beginning of the outbreak.
  As it is seen from parabola, L is prediction of the final size of the epidemic which has been used to fit the logistic curve of the epidemic trajectory. }
  \label{fig:ICCexample}   
\end{figure}

The method described in Section \ref{forecast model} explains that the relationship between the cumulative number of cases and the rate of new cases can be represented as a parabola.
This structure implies that the final size of the epidemic $L_0$, the total number of cases by the end of the outbreak, can be predicted from the domain of this parabola. 
Specifically, when a sample is available from the beginning of an outbreak, we can fit a parabola to the data (plotting the cumulative cases against the rate of new cases) and thereby obtain a predict of the final epidemic size.
Once this final size has been predicted, we then fit a logistic function in the time domain using the predicted final size together with the early-time incidence data. 
This procedure yields forecasts of the cumulative number of cases at each time point (Figure~\ref{fig:ICCexample}).

\noindent As a real example, Supplementary-XXV shows the point prediction procedure for Florida 2023.
Panel (a) shows the data used and panels (b) and (c) shows the fitted parabola and logistic trajectory of the epidemic that can explain the rest of the outbreak in season 2. 

For the general forecast process in Florida, we used data from the outbreak onset up to and including week~$t_0$.
Based on the logistic function fitted to this period, we predicted cumulative number of cases for the subsequent four weeks, namely $t_0+1$, $t_0+2$, $t_0+3$, and $t_0+4$.
In practice, we modeled the cumulative number of cases for week $t$ as a left–censored Poisson random variable.
The Poisson mean $C_{t}$ is given by the ICC curve forecast for week $t$, and support is restricted to $[R_{t-1}, \infty)$, where $R_{t-1}$ is the cumulative number of cases reported to $t-1$.

\noindent Let $N_{t}$ denote the cumulative number of cases for week $t$. 
\noindent We assumed:
\[
N_{t} \sim \text{CPoisson}(C_{t},[0,R_{t-1}))
\]
in which CPoisson$(C_{t},[0,R_{t-1}))$ is a Poisson pmf with parameter $\lambda=C_{t}$ whose domain has been censored on $[0,R_{t-1})$.
Equivalently, the pmf is
\begin{equation}
\Pr(N_{t}=n) =
\begin{cases}
\dfrac{e^{-C_{t}}\,C_{t}^{\,n}/n!}{
1 - \displaystyle\sum_{k=0}^{R_{t-1}-1} e^{-C_{t}}\,\dfrac{C_{t}^{\,k}}{k!}
}, & n \ge R_{t-1}, \\[10pt]
0, & n < R_{t-1}.
\end{cases}
\end{equation}

\subsubsection{Bayesian approach}\label{sec: Bayesian}
Let \(\{X_i\}_{i=1}^{p}\) denote the weekly number of reported dengue cases in \(p\) weeks of an outbreak.  
Based on the theory of the ICC curve for dengue, the temporal trajectory of the outbreak is represented by a logistic random variable \(T\).  
In practice, however, surveillance data are subject to both under and over reporting, and the final size parameter \(L\) can be misestimated when a parabolic approximation is fitted directly to the observations.  
To accommodate these imperfections, we embed the reporting error in a Bayesian model, treating it as an observational noise term so that the associated uncertainty is approximated through to the posterior inference.

\subsubsection{Trajectory of the outbreak mechanism}

As we saw in section \ref{forecast model}, the outbreak trajectory is described by a logistic random variable.
\begin{equation}
  \mathbf{T} \;\sim\;
  \operatorname{Logistic}\!\bigl(\mu(\mathbf{O}_{p},L),\,\delta(\mathbf{O}_{p},L)\bigr),
  \label{eq: likelihood}
\end{equation}

\noindent where
$\mathbf{O_p}=(O_1,\dots ,O_{p})^{\!\top}$ and $L$ are the latent vector of true weekly cases and the final size of the epidemic, respectively.
In addition, $\mu(\mathbf{O_p},L)$ and $\delta(\mathbf{O_p},L)$ are the location and scale parameters obtained from the ICC curve. 

\subsubsection{Observation model}\label{sec: observation model}
Each weekly report is modeled as the true incidence perturbed by additive Gaussian measurement error. 
Denote the number of new cases to week~\(p\) by
\[
\mathbf{x}_{p}=(x_{1},\ldots,x_{p})^{\top},
\qquad
\mathbf{O}_{p}=(O_{1},\ldots,O_{p})^{\top},
\]
where \(\mathbf{O}_{p}\) is the corresponding latent (error–free) vector.

\noindent We assume
\begin{equation}
  \mathbf{O}_{p} \;=\; \mathbf{x}_{p} + \mathbf{e}_{p},
  \qquad
  \mathbf{e}_{p} \sim \mathcal{N}\!\bigl(\mathbf{0},\,\delta_{e}\mathbf{I}_{p}\bigr)
  \;\Longrightarrow\;
  \mathbf{O}_{p} \mid \mathbf{x}_{p} \sim
  \mathcal{N}\!\bigl(\mathbf{x}_{p},\,\delta_{e}\mathbf{I}_{p}\bigr).
\end{equation}

\noindent Because weekly case numbers are intrinsically nonnegative, we imposed a
left‑truncated (censored) Gaussian prior on every component of the
\begin{equation}
  \mathbf{O}_{p}  \mid \mathbf{x}_{p} \;\sim\;
  \operatorname{TN}_{[0,\infty)}\!\bigl(\mathbf{x}_{p},\,\delta_{e}\mathbf{I}_{p}\bigr),
  \label{eq:obs_model}
\end{equation}
where \(\operatorname{TN}_{[0,\infty)}(\mu,\Sigma)\) denotes the multivariate
normal distribution \(\mathcal{N}(\mu,\Sigma)\) restricted to the non‑negative
orthant.  
This specification guaranties \(O_{t}\ge 0\) for every week
\(t\) while retaining the familiar Gaussian shape within the admissible region.
The value of \(\delta_{e}\) should reflect the context: specifically, the precision of the reporting and diagnostic protocols used. 
Consequently, \(\delta_{e}\) will be higher (allowing greater observation noise) in settings where testing is costly or infeasible and case confirmation is based primarily on clinical symptoms, and lower in settings with standardized laboratory confirmation (e.g. NAAT / RT-PCR, NS1, or repeated serology) and rigorous reporting procedures.
For this case study, in this setting, we assume that the weekly reported data of new cases are accurate.
This assumption is supported by the rigorous evaluation of suspected dengue patients in Florida and the use of standardized diagnostic protocols for individuals referred to hospitals or public health laboratories \cite{FDOH_BPHL_DENV_PCR_2023,CDC_Molecular_Tests_Dengue_2025,CDC_Clinical_Testing_Dengue_2025}.
Consistent with this high ascertainment, we placed a tight observation prior to the reported counts, specifying a variance of \(0.3\) for \(\pi(\mathbf{O})\).
This corresponds to a standard deviation of approximately \(0.55\) cases, which reflects small, but nonzero, week-to-week reporting noise expected under aggregated laboratory-confirmed surveillance.
In practical terms, this prior allows draws that fall modestly below or above the reported count while strongly concentrating mass near the observed value, thereby encoding reporting precision without treating counts as error-free. 

\subsubsection{Model of final size of the epidemic}
We assigned a left-truncated Gaussian prior to the final size parameter \(L\).  
The mean of the distribution (\(L_{0}\)) is obtained from the ICC curve method to weekly observations \(\{X_{i}\}_{i=1}^{p}\), and the prior variance is denoted by \(\delta_{L}\).  
To ensure that the final size cannot be less than the cumulative number of reported cases \(S=\sum_{i=1}^{p}X_{i}\), the distribution is truncated below at \(S\):
\begin{equation}
  L \;\sim\;
  \mathcal{N}_{\,[S,\,\infty)}\!\bigl(L_{0},\,\delta_{L}\bigr),
  \label{eq:L_model}
\end{equation}
For settings where no prior information on \(\delta_{L}\) is available, we adopted a very large variance prior for \(L\) that approximates a non-informative prior, which is appropriate when no prior knowledge is imposed.



\subsubsection{Posterior distribution}
Applying Bayes’ theorem with the parameter vector
\(\theta = (\mathbf{O},\,L)\) yields
\[
\begin{aligned}
\pi\!\bigl(\theta \mid T = t\bigr)
      &\;\propto\; f\!\bigl(T \mid \theta\bigr)\,f(\theta) \\[6pt]
      &= f\!\bigl(T \mid \mathbf{O},L\bigr)\,
         f\!\bigl(\mathbf{O},L\bigr) 
      \;\propto\; f\!\bigl(T \mid \mathbf{O},L\bigr)\,
         f\!\bigl(L \mid \mathbf{O}\bigr)\,
         \pi(\mathbf{O}).
\end{aligned}
\]
To facilitate computation, we assumed the prior independence between
\(L\) and \(\mathbf{O}\).  
Under this assumption, the joint posterior simplifies to
\[
\pi\!\bigl(\mathbf{O},L \mid T = t\bigr)
      \;\propto\;
      f\!\bigl(T \mid \mathbf{O},L\bigr)\,
      f(L)\,
      \pi(\mathbf{O}).
\]
Combining the truncated Gaussian priors (\ref{eq:obs_model}) and (\ref{eq:L_model}) with the logistic
likelihood (\ref{eq: likelihood}) yields the unnormalized posterior density.
\begin{equation}
\begin{aligned}
\pi(\mathbf{O}_{p},L \mid T=t)
&\;\propto\;
\frac{%
  \delta(\mathbf{o}_{p},L)\,
  \exp\!\bigl[-\delta(\mathbf{o}_{p},L)\,
               \bigl(t-\mu(\mathbf{o}_{p},L)\bigr)\bigr]}{%
  \bigl[1+\exp\!\bigl(-\delta(\mathbf{o}_{p},L)\,
                     \bigl(t-\mu(\mathbf{o}_{p},L)\bigr)\bigr)\bigr]^{2}}
\\[6pt]
&\;\times\;
\mathbf 1_{\{\mathbf{o}_{p}\ge 0\}}\,
(2\pi\delta_{e})^{-p/2}\,
\exp\!\Bigl[-\tfrac{1}{2\delta_{e}}\,
            \lVert\mathbf{o}_{p}-\mathbf{x}_{p}\rVert^{2}\Bigr]
\\[6pt]
&\;\times\;
\mathbf 1_{\{L\ge S\}}\,
(2\pi\delta_{L})^{-1/2}\,
\exp\!\Bigl[-\tfrac{(L-L_{0})^{2}}{2\delta_{L}}\Bigr].
\end{aligned}
\label{eq:posterior_reduced}
\end{equation}
Here, \(\mathbf{1}_{\{\mathbf{o}_{p}\ge 0\}}\) and \(\mathbf{1}_{\{L\ge S\}}\) enforce the non-negative support.
It is important to note that what we have done is essentially to use the information from $p$ weeks to obtain a posterior as a function of time $t > p$.
This approach enables us to infer the future trajectory on the basis of the available data.
\subsubsection{Sampling and prediction}
The unnormalized posterior in equation~\eqref{eq:posterior_reduced} lacks a
closed‐form normalizing constant; consequently, we employ Markov–chain
Monte Carlo (MCMC) to draw samples from the joint distribution
\(\bigl(\mathbf{O}_{p},L\bigr)\mid T=t\).
At each iteration, a proposal \(\bigl(\mathbf{O}_{p}^{\ast},L^{\ast}\bigr)\)
is obtained by adding independent Gaussian increments to the current state and
reflecting the result in the admissible regions
\(\mathbf{O}_{p}\ge\mathbf 0\) and \(L\ge S\).
The candidate is accepted with the Metropolis–Hastings probability computed
from the posterior kernel in~\eqref{eq:posterior_reduced}; otherwise, the chain
retains its previous value.
Because every smooth functional of
\(\bigl(\mathbf{O}_{p},L\bigr)\) inherits a posterior distribution by
transformation of retained draws, no further numerical integration is
required.  
In particular, after each accepted update we apply the ICC‐curve procedure to
\(\bigl(\mathbf{O}_{p},L\bigr)\) to obtain the logistic parameters
\(\mu\bigl(\mathbf{O}_{p},L\bigr)\) and \(\delta\bigl(\mathbf{O}_{p},L\bigr)\).
The predictive cumulative number of cases at time \(t\) is then evaluated as
\[
  C_{t} \;=\;
  \frac{L}{
        1+\exp\!\Bigl\{-\delta\bigl(\mathbf{O}_{p},L\bigr)
                       \bigl[t-\mu\bigl(\mathbf{O}_{p},L\bigr)\bigr]\Bigr\}},
\]
yields a posterior sample \(\{C_{t}\}\) at no additional computational
cost.  
A histogram or kernel density estimate of this sample approximates the
predictive density
\(\pi\!\bigl(C_{t}\mid\mathbf{T}=\mathbf{t}\bigr)\), providing a fully Bayesian
forecast that propagates both epidemiological uncertainty and measurement
error.
A detailed description of this MCMC algorithm is provided in Appendix C.

\subsection{Evaluation}
In this section, we present the methodologies used to evaluate the results of our work, encompassing point estimation, a Bayesian inference framework, and diagnostics for Markov chain Monte Carlo (MCMC) sampling. To address real-time applicability, we evaluated the models in a rolling forecasting setting that emulates operational use: at each epidemiological week, we fit the model using only the data available up to that week, generated predictions for the subsequent four weeks, and then compared those forecasts with the subsequently reported observations; we then advanced the training window by one week and repeated this procedure throughout the 2022–2023 seasons.

\subsubsection{Point prediction}
To evaluate the performance of the model, we used three complementary approaches to assess its precision.
For point prediction, we calculated the root mean square error (RMSE) between point prediction and the real value of cumulative cases reported for each week.
As another metric to evaluate the accuracy of the point prediction, the distribution of absolute value of error (AVE) has also been calculated. 
In addition, we evaluated point predictions using the logarithmic score adopted by CDC in its open prediction challenge for the West Nile virus (WNV) \cite{holcomb2023evaluation}. 
This method evaluates the point predictions based on their uncertainty, rather than only the single point estimate.
If the observed number of cases falls in a region of the probability density function of the point prediction with higher probability, it receives a higher score.
Details are provided in Appendix D, and Figure~\ref{fig: evaluationprocess} illustrates the evaluation and scoring process.

\subsubsection{Bayesian approach}
For Bayesian point prediction, we have considered the median and mean of the posterior as Bayesian point prediction and based on that calculated $RMSE_{B}$, which is RMSE between Bayesian prediction and the value of the real cumulative cases reported for weeks.
We also used the logarithmic score for the Bayesian approach.

\begin{figure*}[htbp]
    \centering

    \begin{subfigure}{1\linewidth}
        \centering
        \includegraphics[height=0.13\textheight]{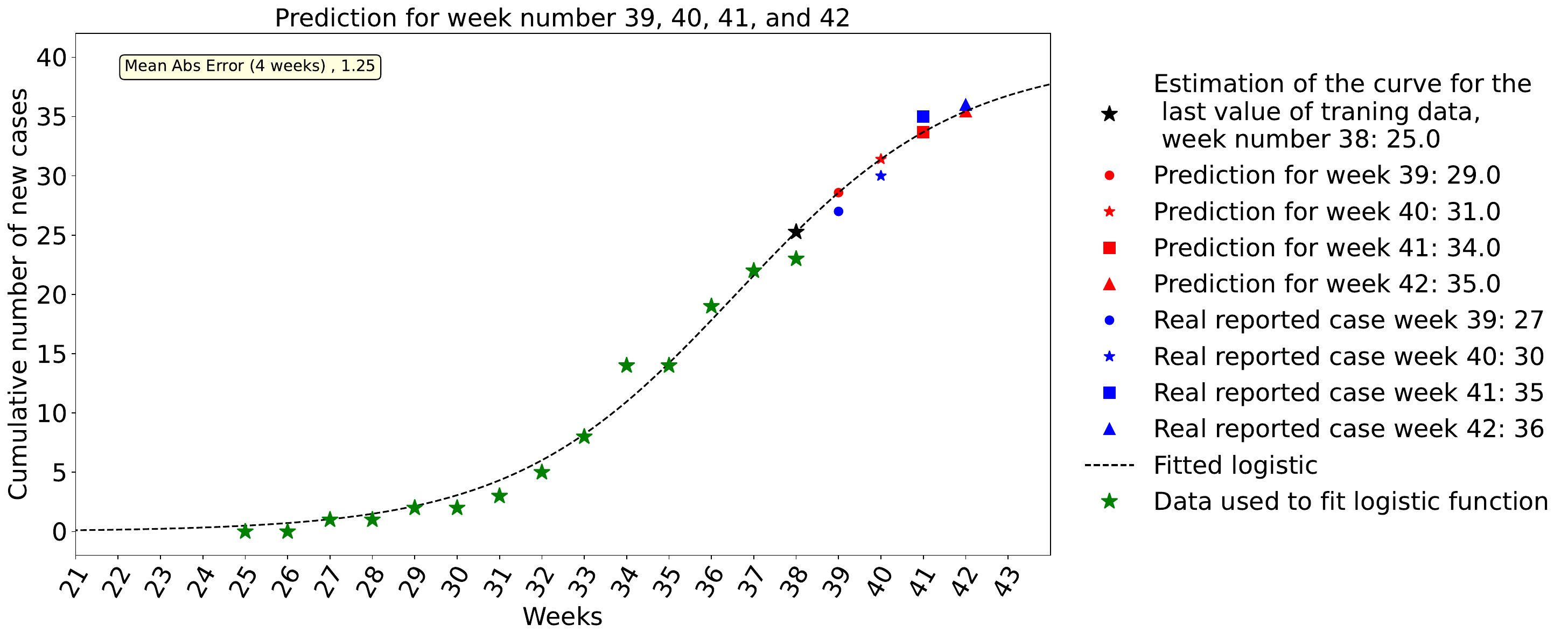}
        \caption{}
        \label{fig:logistic}
    \end{subfigure}
    \hfill
    \begin{subfigure}{1\linewidth}
        \centering
        \includegraphics[height=0.24\textheight]{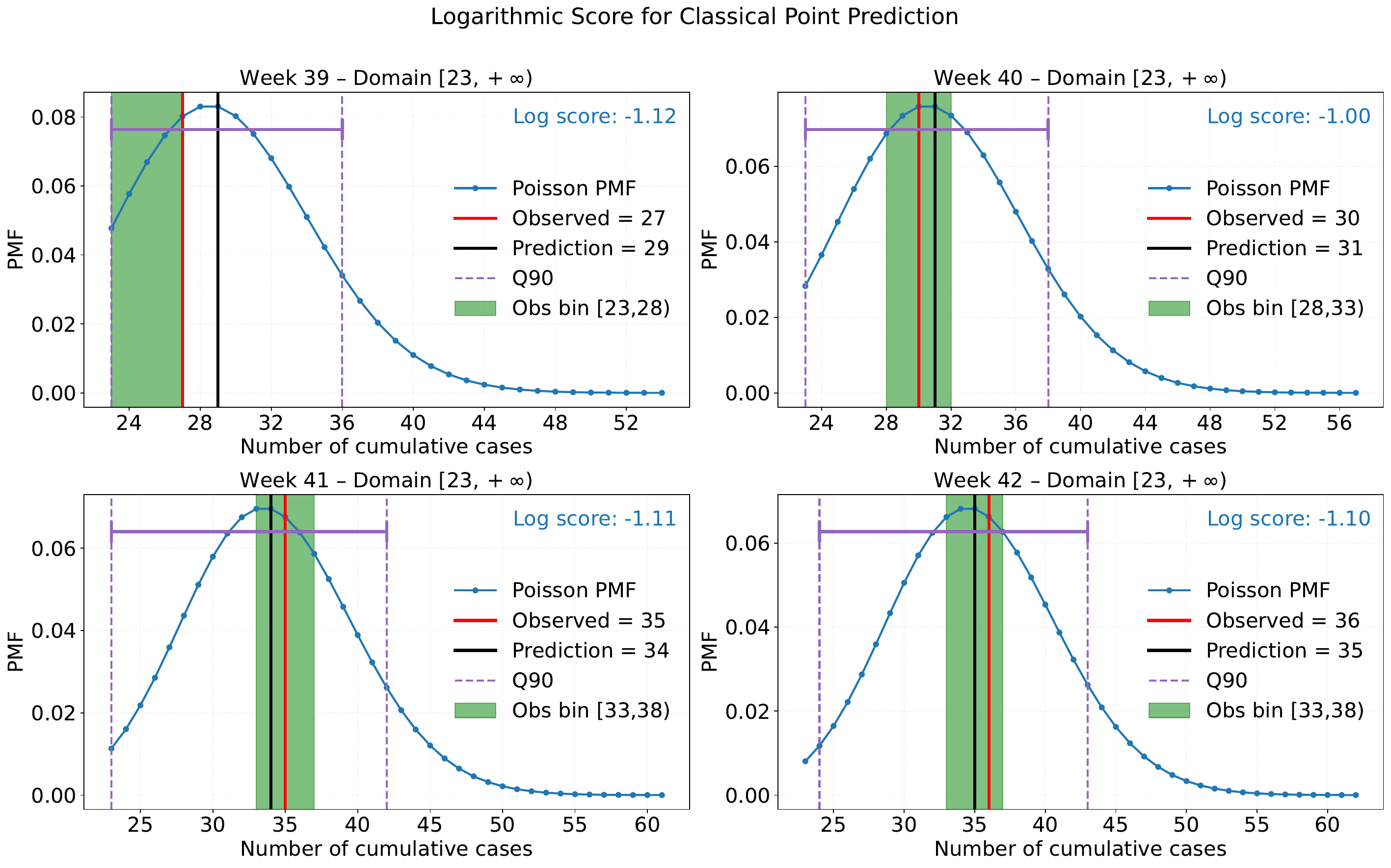}
        \caption{}
        \label{fig:poisson}
    \end{subfigure}

    \caption{(a) Shows the logistic trajectory of the epidemic obtained by ICC-curve method and the values of prediction for four weeks ahead and the real reported cases; (b) Shows the result of the prediction based on the censored Poisson model. The prediction obtained in panel (a) is used as rate for the Poisson model. Also for each week winner epidemiological bins has been shown by orange region}
    \label{fig: evaluationprocess}
\end{figure*}

\subsubsection*{2.5.3. MCMC diagnostics}
In addition to evaluating predictive performance, we assessed the quality of the MCMC sampling itself. 
We ran four independent Metropolis--Hastings chains, each long enough to yield a sufficiently large effective sample size (ESS).
For each chain, the first \(1\times 10^{4}\) accepted draws were discarded as burn--in to remove transient behavior, and the remaining samples were thinned by retaining every fifth draw to reduce autocorrelation.
Convergence and mixing were evaluated using three standard diagnostics: (i) the split Gelman--Rubin statistic, \(\hat{R}\), to assess between--chain convergence; (ii) Geweke's early--versus--late \(Z\)-score to detect residual nonstationarity within chains; and (iii) the integrated-autocorrelation-based ESS, reported for each chain individually and for the pooled set of chains.

\section{Results} \label{sec: results}
In the following sections, we present the results of the ICC model, and the evaluation of its performance.
We also summarize the output of the Bayesian model, including the MCMC diagnostic checks and the predictive results.
Based on the time series of the number of new cases , four distinct outbreak periods were considered in Florida between 2022 and 2023.
The time series inspection (Figure \ref{fig:Florida 2022-23-seasons}) indicates that
the first outbreak occurred between week 13 and week 21. 
The second outbreak lasted 43 weeks, beginning in week 25 and ending in week 67. 
The third outbreak spanned 11 weeks, from week 77 to week 87.
The final outbreak, which was the most severe in terms of the number of reported cases, began in week 88 and persisted until week 103, with a total duration of 16 weeks.

\subsection{Point prediction}
As explained in Section \ref{sec: data}, to demonstrate the method in practice, we collected data (2022-2023) on new locally acquired cases in Florida from the US Centers for Disease Control and Prevention (CDC).
In total, 258 weekly forecasts were generated, comprising 69 one-week ahead predictions, 66 two-week ahead predictions, 63 three-week ahead predictions, and 60 four-week ahead predictions across all outbreaks.

\begin{figure*}
    \centering
    \includegraphics[width=0.7\linewidth]{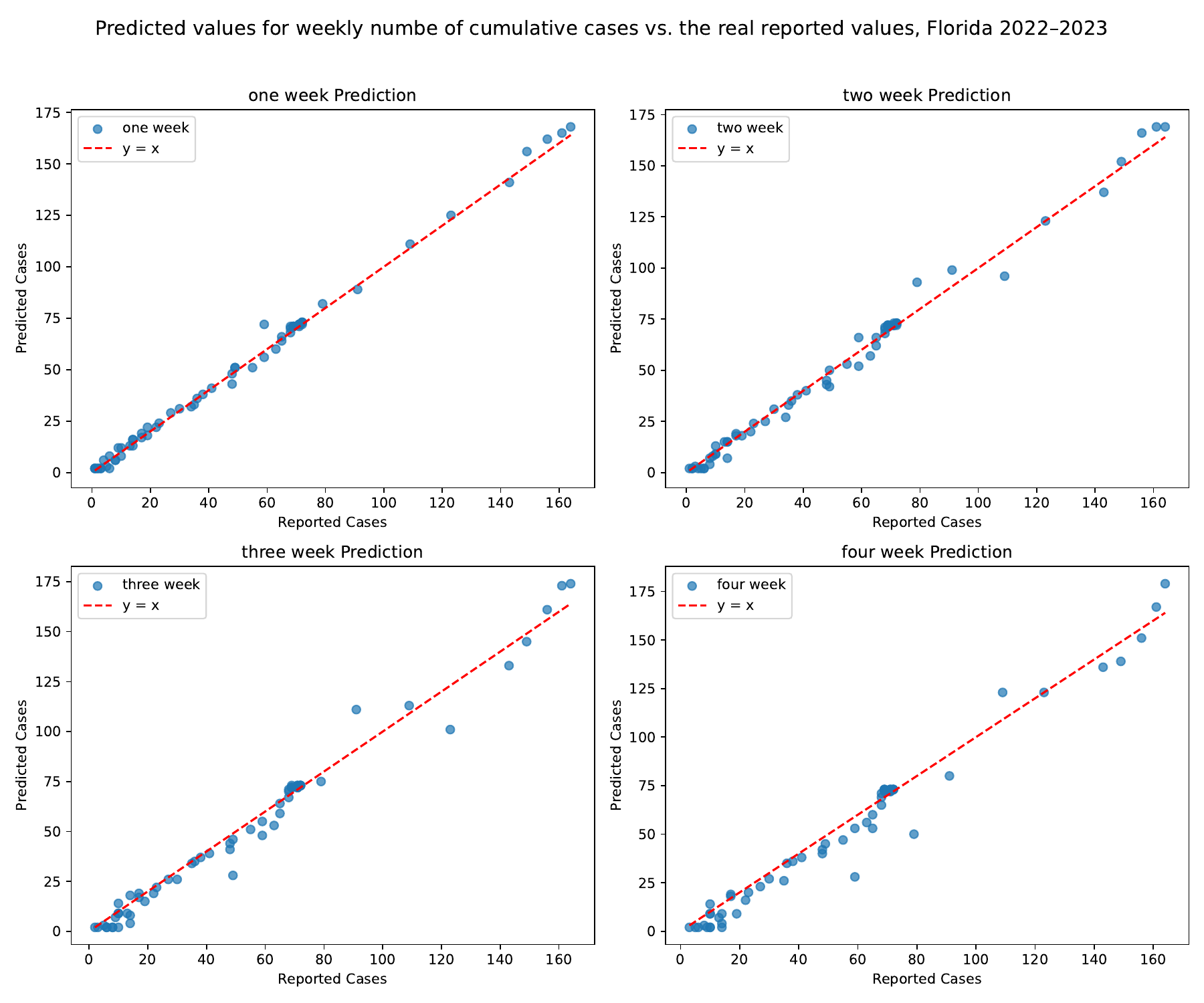}
    \caption{One– to four–week ahead predictions of weekly cumulative cases versus reported values in Florida (2022–2023). Each panel corresponds to a forecast horizon (1, 2, 3, or 4 weeks). Reported cases are on the horizontal axis; predicted cases are on the vertical axis. The red dashed line
}
    \label{fig: prediction_vs_real}
\end{figure*}

The distribution of AVE shows that for a one-week ahead prediction, 43.48\% of the prediction's AVE were less than or equal by one and 78.26\% of the prediction's AVE were less than or equal by two cases.
More than 95\% of the predictions had errors of five cases or fewer, while only less than 4.35\% of the predictions have errors greater than or equal to 6.
For the forecasts for two-weeks ahead, approximately $10.61\%$ of the predictions did not show an error in case counts, while nearly half ($43.94\%$) deviated by at most one case.
Errors in two or fewer cases accounted for $59.1\%$ of the predictions, and $80.3\%$ were within five cases.
In addition, $16\%$ of the predictions had errors greater than or equal to 6 cases.  
For the three-week forecasts, exact matches between the predicted and observed cases occurred in $3.2\%$ of the predictions.
Deviations in one or two cases represented a little less than half of the results ($30.16\%$). 
The cumulative accuracy reached $74.6\%$ for errors of up to five cases and exceeded $92\%$ within ten cases.  
For the four-week ahead forecasts $21\%$ of the predictions, they deviated by only one case.
More than half of the forecasts ($53.33\%$) were within four cases. 
The cumulative accuracy exceeded $93.33\%$ for deviations up to 12 cases.
Figure \ref{fig: prediction_vs_real} shows the prediction versus real values for the weekly number of cumulative cases. 
The general RMSE for all outbreaks was $5.7$, calculated over 258 weekly predictions encompassing forecast horizons of one, two, three, and four-weeks ahead.
Across all outbreaks, the RMSE by prediction horizon was $3.81$ for one--week, $3.57$ for two--week ahead, $2.99$ for three--week ahead and $10.7$ for predictions for four--week ahead.
All results, with additional details, are provided in Supplementary XI.

\subsubsection{First outbreak} 
To predict the number of new cases at the beginning of 2022, we used data from the first outbreak, starting from its onset (week 13) until the end of outbreak week 21 of 2022 (Figure \ref{fig:Florida 2022-23-seasons}).
In the first step of fitting the parabolic and logistic functions, we considered data from three consecutive weeks, namely weeks 13, 14, and 15, which reported the case 0, 0, and 1, respectively. 
The predictions were then generated for cumulative cases on horizons of 1, 2, 3, and 4 weeks ahead, corresponding to weeks 16, 17, 18, and 19.
In each step, we add one more week to the data for fitting purposes and predicted the next four weeks values.
The predicted value for cumulative cases was rounded to the closest integer number (Supplementary XII). 
All the process of fitting and prediction has been plotted in Supplementary I.

Across outbreak~1, for one--week ahead predictions we observed only three error values (1, 2, and 4): 33.3\% of forecasts had an error of exactly one case, 83.3\% differed from the observed cases by at most two, and all errors were bounded by four.
For two--week ahead predictions, we again observed only three error values (1, 3, and 4): 50.0\% of the forecasts had an error of exactly one case, 66.7\% were within three cases, and all errors were no greater than four.
For three--week ahead predictions, we observed three error values (1, 4, and 8): 33.3\% of the forecasts had an error of exactly one case, 83.3\% were within four cases, and all errors were bounded by eight.
At the four--week ahead horizon, we observed three error values (1, 4, and 8), with 33.3\% of forecasts differing from the observed count by only one case, 66.7\% within four cases, and all errors were capped at eight (Supplementary XIII).
For outbreak 1, the RMSE values were 2.24 for the one--week ahead predictions, 2.71 for the two--week ahead predictions, 4.36 for the three--week ahead predictions, and 5.20 for the four--week ahead predictions.

\subsubsection{Second outbreak}
For the second outbreak, in the first step we used data from weeks 25, 26, and 27 to forecast weeks 28, 29, 30 and 31, and in each step we added one week to our training data and did a prediction of 1, 2,3--, and 4-weeks ahead, as explained before.
The predictions results for 1, 2, 3-- and 4--week ahead have been published in Supplementary XIV and all the plots related to Poisson PMF and the fitted parabola and logistic function have been attached to Supplementary II.
Across outbreak 2, for one week ahead forecasts, 24.39\% of the predictions exactly matched the observed values, while 60.98\% deviated by at most one case and 87.8\% by no more than two. 
Errors up to three cases accounted for 95.12\% of the forecasts, and all were bounded within five.
Two weeks ahead, 15\% of the predictions had zero error, more than half (52. 5\%) were within one case, and approximately three quarters (70\%) remained within two. 
82.5\% of the forecasts had errors less than or equal to three captured and all deviations were limited to seven cases.
The three-week horizon showed that 41.03\% were within one and 63.85\% stayed within two.
By three errors, nearly three-quarters (69.23\%) were included and all deviations were capped at ten.
For the four week prediction, 52.63\% of the forecasts had errors of no more than three, 78.95\% within six, and all errors were restricted to twelve (Supplementary XV).
For outbreak 2, the RMSE values were $1.76$ for the predictions made one week in advance, $2.65$ for two weeks, $3.73$ for three weeks, and $5.31$ for four weeks.

\subsubsection{Third outbreak}
For the initial prediction window, we used data from weeks 77, 78, and 79 to forecast weeks 80, 81, 82, and 83.
In total, this setup produced 8, 7, 6, and 5 predictions in the one--, two--, three--, and four--week ahead horizons, respectively, and the results are shown in Supplementary XVI and Supplementary III.
For outbreak three, the one--week ahead forecasts showed that $25\%$ of the predictions were exact (error zero), $37.5\%$ were within one case and all predictions ($100\%$) were within three cases.
At two weeks ahead, $57.14.3\%$ of the forecasts had zero error, $100\%$ were within two cases.
For the three--week ahead predictions, the absolute error took only four values (0, 2, 4, and 6); $16.7\%$ of the predictions were exact, $50.0\%$ were within two cases and $83.3\%$ were within four cases of the observed counts.
For the four--week ahead predictions, the absolute error took five values (1, 2, 5, 6, and 7); none of the predictions were exact, but 40.0\% were within two cases, 60.0\% were within five cases, 80.0\% were within six cases, and all predictions were within seven cases of the observed counts
(Supplementary XVII).
For outbreak three, the RMSE values by prediction horizon were as follows: one--week ahead predictions had an RMSE of $1.80$, two-week-ahead predictions $1.51$, three-week ahead predictions $3.55$, and four-week ahead predictions $4.79$.

\subsubsection{Fourth outbreak}
The fourth outbreak began in week 87 and continued for 16 weeks. 
Supplementary XVIII presents the sets of 14, 13, 12 and 11 weekly forecasts of cumulative case counts, together with the corresponding observations, for the one--, two--, three--, and four--week ahead prediction horizons, respectively.
In the one--week ahead forecasts, $42.86\%$ of the predictions had an absolute error of 2 cases.
Overall, $57.14\%$ of the forecasts were within 3 cases of the observed counts, and approximately $80\%$ were within 5 cases.
Detailed results of the distribution of the AVE for the one-, two--, three--, and four--week ahead predictions are provided in Supplementary XI.
The root mean square error (RMSE) for the one-week ahead forecast is $\mathrm{RMSE}=5.02$. 
For the two-week ahead forecast, the RMSE increases slightly to $\mathrm{RMSE}=8.07$. 
At three-week ahead, the RMSE rises more notably to $\mathrm{RMSE}=12.62$, 
and by four-week ahead, it reaches $\mathrm{RMSE}=15.66$.

\begin{figure}
    \centering
    \includegraphics[width=0.7\linewidth]{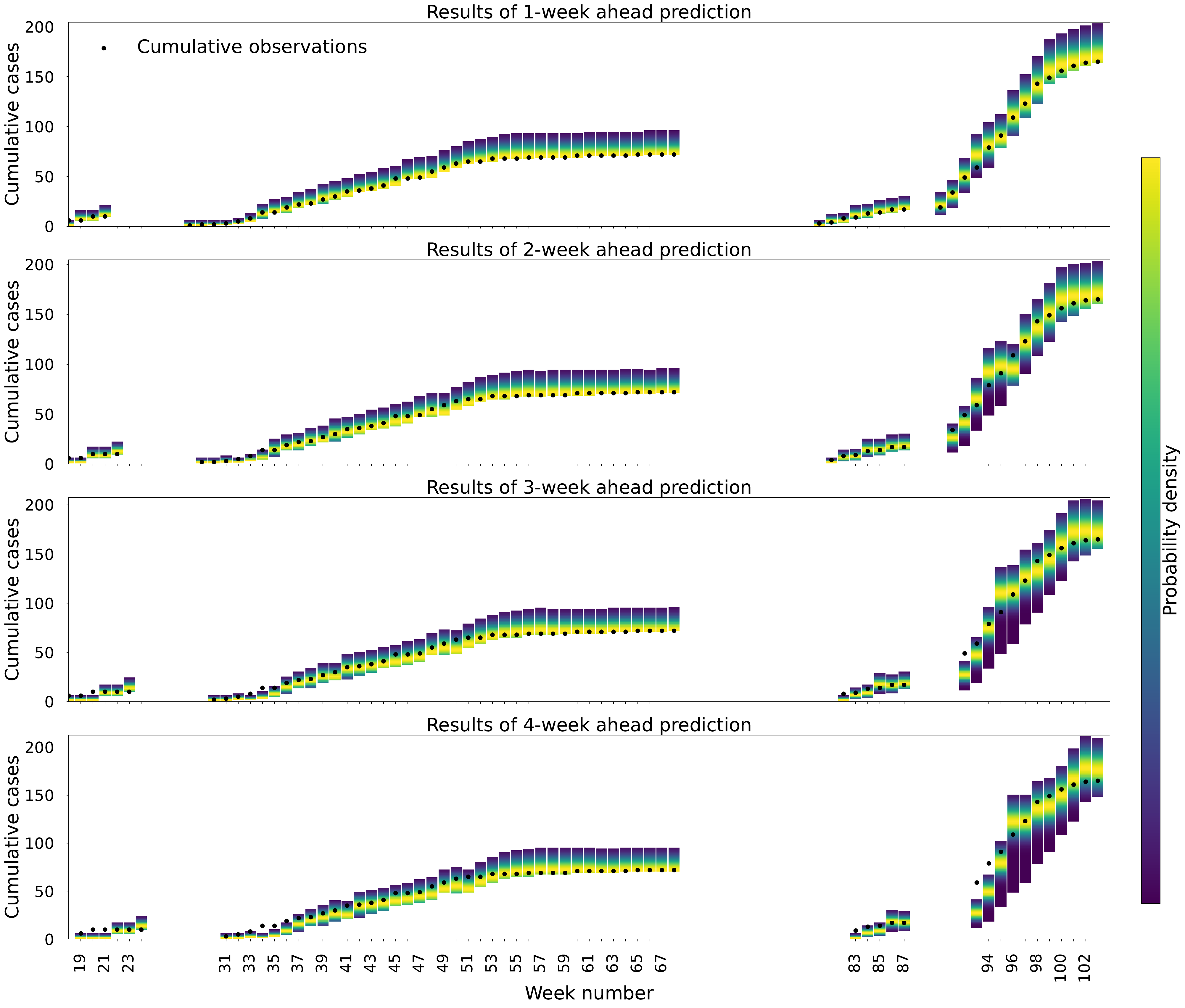}
    \caption{Forecasts of cumulative cases by week and their probability densities for four Florida outbreaks (2022–2023) at 1–4-week horizons. Black dots mark observed cases; lighter shading denotes higher probability density.}
    \label{fig:summary_all}
\end{figure}

As a summary of the results, all predictions for one--, two--, three-- and four-weeks ahead of the cumulative weekly number of cases for all four outbreaks (2022–2023), together with their censored Poisson predictive distributions, are shown in Figure \ref{fig:summary_all}.
The figure overlays the observed cumulative case counts with the weekly predictive distributions, allowing visual comparison between predicted and observed values across forecast horizons and outbreak periods. 
All details of these plots and their weekly distributions have been provided in Supplementary I--IV.

\subsection{Bayesian model}
\subsubsection{MCMC diagnostics}
\begin{figure*}[htbp]
    \centering
    
    \begin{subfigure}{0.45\linewidth}
        \centering
        \includegraphics[width=\linewidth]{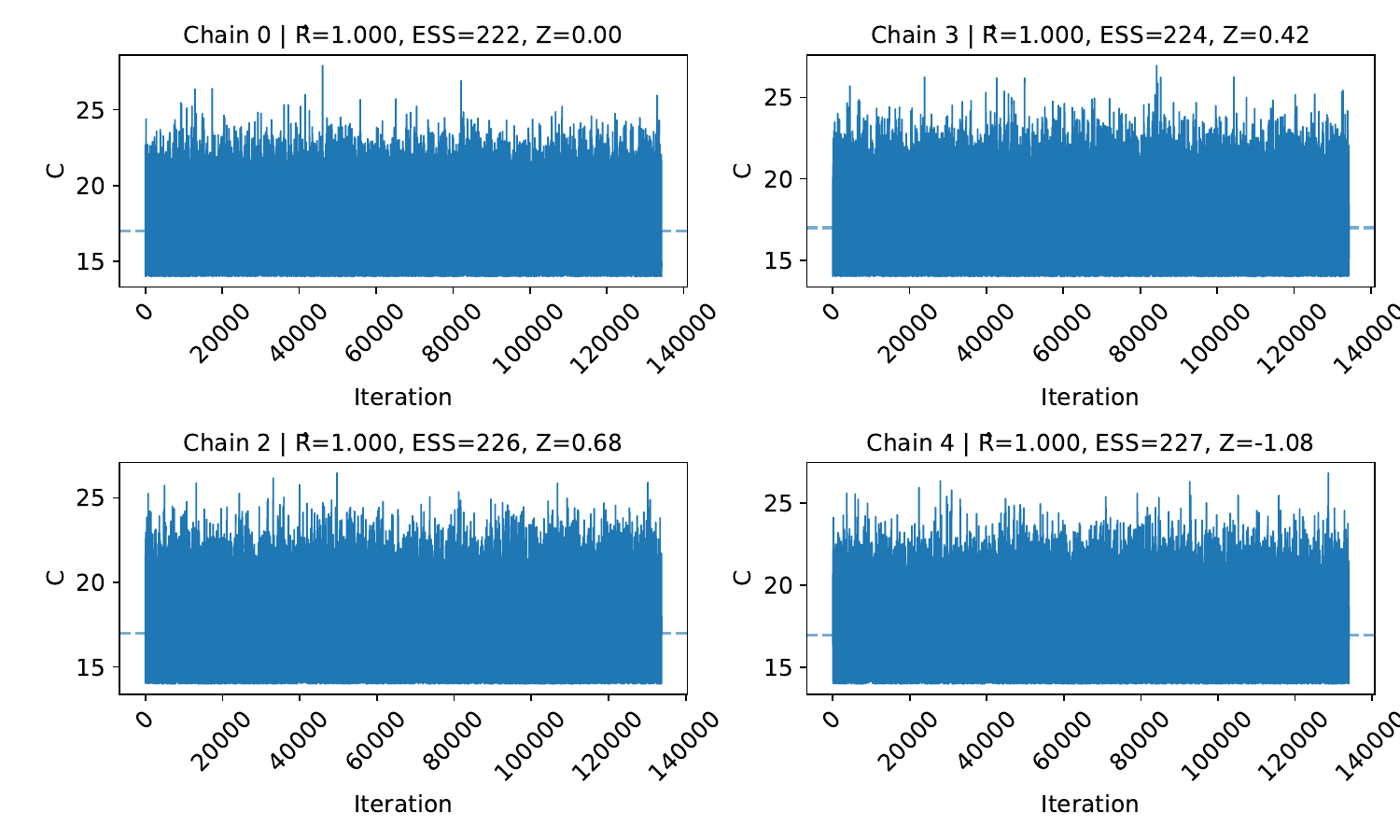}
        \caption{}
        \label{fig:pdf}
    \end{subfigure}
    \hfill
    \begin{subfigure}{0.5\linewidth}
        \centering
        \includegraphics[width=\linewidth]{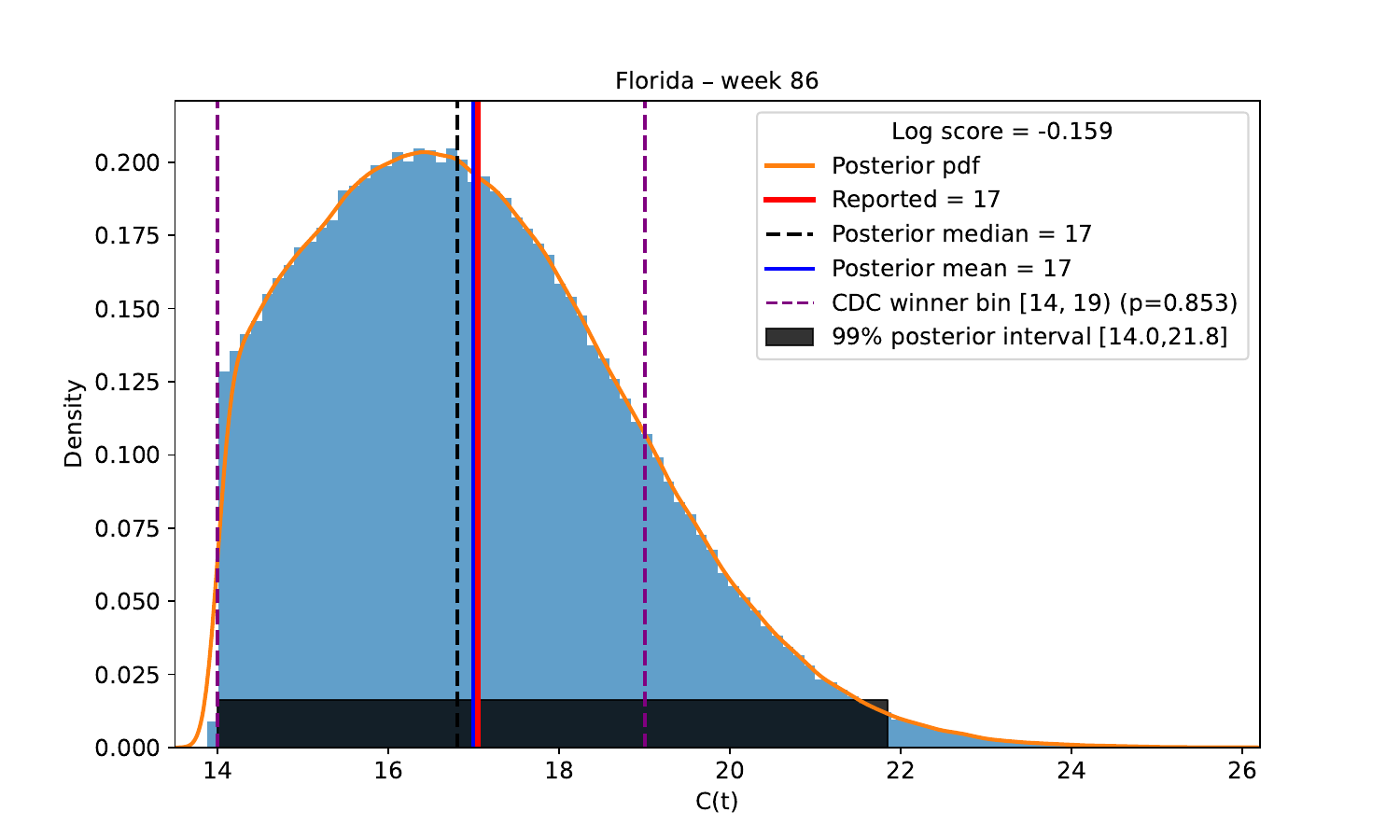}
        \caption{}
        \label{fig:week}
    \end{subfigure}

    \caption{(a) shows four MCMC chains of MCMC we have considered for plotting the posterior. 
    (b) shows the posterior PDF, Bayesian predictions, and the real observation for cumulative number of cases for week 86.}
    \label{fig:all_three}
\end{figure*}

To sample from the posterior distribution, we ran four independent MCMC chains. 
Convergence and mixing were evaluated using standard diagnostics, applied both to the posterior samples of the weekly cumulative number of cases and to the posterior samples of the weekly observations.  
Figure~(\ref{fig:all_three}) shows the resulting trace plots together with their corresponding diagnostic statistics for week 86.
After discarding burn-in and applying thinning, each chain achieved a sufficiently effective sample size (ESS).
The Geweke Z scores satisfied $|Z|<2$ for all chains, and the Gelman--Rubin statistic was $\hat{R}\approx 1.00$ throughout. 
Visual inspection of the trace plots showed stable behavior with no discernible trends or drifts.
The diagnostic results for each chain are reported directly on the corresponding MCMC trace plots in the Supplementary Bayesian materials (Supplementary V-VIII).

\subsubsection{Bayesian predictions}
During the first outbreak, weeks 16-21, Bayesian point forecasts increased from 3 to 13 cases, with medians that matched the means in all weeks except week 21 (median 12).
Mean of absolute errors 2 shows a small error for the weekly Bayesian prediction.
For the first outbreak, $RMSE_B$ is 2.27 for the Bayesian mean forecasts and 2.08 for the Bayesian median forecasts, since $RMSE_B$ can be computed separately for each type of point prediction; notably, $RMSE_B$ for the Bayesian median (2.08) is smaller than the corresponding value for the censored Poisson model (2.24).
The Bayesian model achieved logarithmic scores lower than the censored Poisson model in every week except the first two weeks of the outbreak.
All details of the prediction , errors, and scores have been provided in Supplementary V and XX.

During the second outbreak (weeks 29–66), Bayesian point forecasts increased from 3 to 85 cases and generally followed the weekly cumulative reported counts. 
Absolute error values were small through the middle of the outbreak (mostly 0–3 cases up to week 55) and increased near the late plateau (reaching 12–13 cases by weeks 65–66).
The Bayesian mean absolute error was 3.71 and 3.79 for the posterior mean and median forecasts, respectively, and the $RMSE_{B}$ was 5.23 and 5.35 for the Bayesian mean and median. 
In 23 of the 38 weeks, the Bayesian log scores were closer to zero than the corespondent censored Poisson model (Supplementary XXI and VI).

For the third outbreak, from weeks 80–87, Bayesian point forecasts increased from 3 to 20 cases, while reported cases increased from 3 to 17. 
The errors were small throughout (0–3 cases), with an exact match in week 86 and near-matches in weeks 80, 81, 84, and 85. 
The mean absolute error was 1.50 and 1.25 for the posterior mean and median forecasts, respectively, and the corresponding Bayesian RMSE values were 1.73 and 1.58—both markedly lower than the corresponding RMSE of 4.36 from the censored Poisson model. 
Probabilistically, the Bayesian model outperformed the Poisson benchmark in 7 of 8 weeks: its log scores were less negative in weeks 81–87 (Supplementary XXII and VII).

\begin{figure}
    \centering
    \includegraphics[width=1\linewidth]{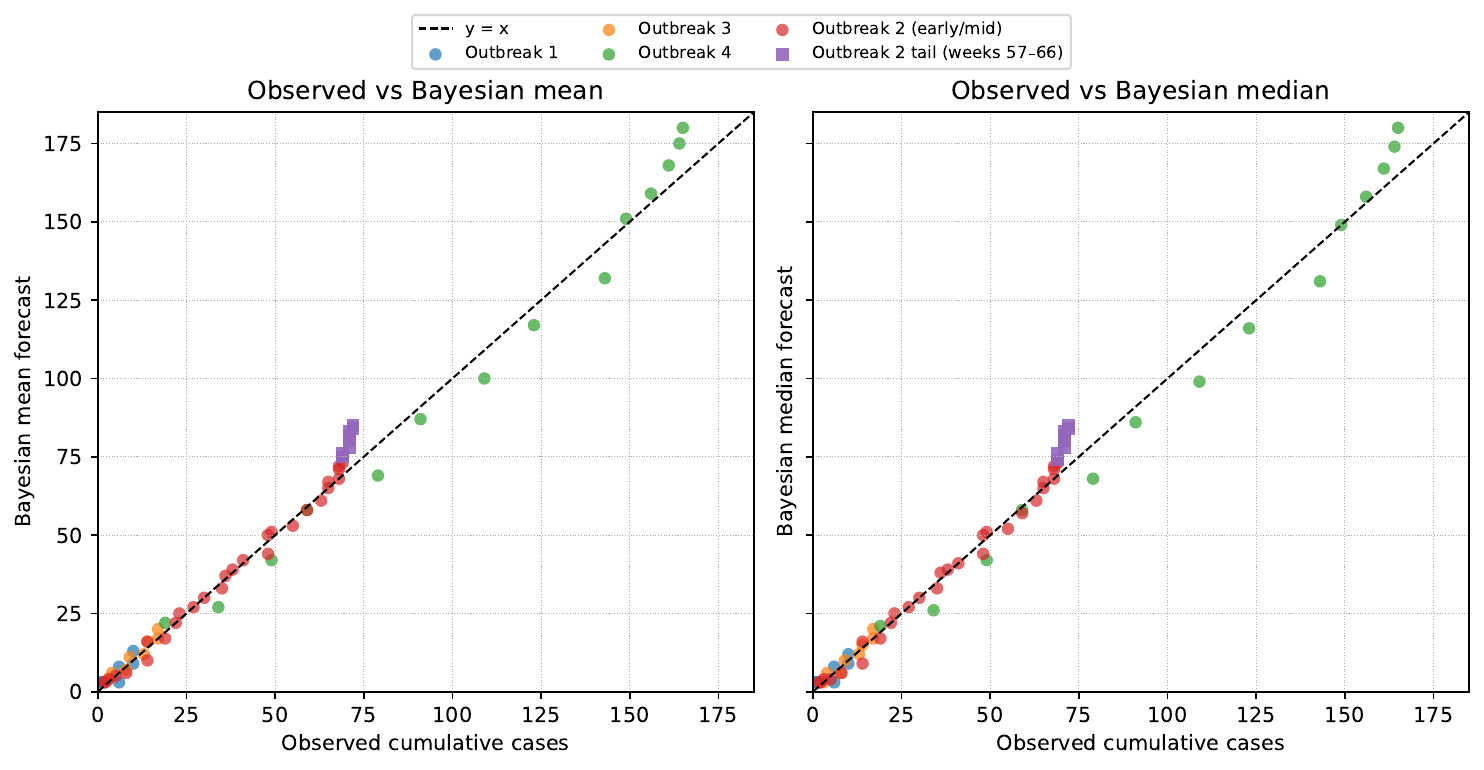}
    \caption{Left panel shows predicted versus observed cumulative number of cases for the four outbreaks in 2022–2023 using the Bayesian mean as the point prediction, while the right panel shows the same comparison using the Bayesian median as the point prediction.}
    \label{fig: bayesianx=y}
\end{figure}

During the fourth outbreak (weeks 90–103), Bayesian point forecasts increased from 22 to 180 cases, with medians closely tracking the means (21 to 180), while reported counts rose from 19 to 165. 
AVEs ranged from 0 to 15 cases and were generally moderate, with small discrepancies (2–4 cases) in weeks 90, 93, 95, and 100 and an exact match in week 99.
Probabilistically, the Bayesian model outperformed the Poisson benchmark in 6 of 14 weeks, mainly during the mid-outbreak period and early plateau (weeks 93, 95 and 99–102), while the Poisson model achieved slightly higher log scores during weeks with sharper incidence changes (e.g., 91–92, 94, 96-98 and 103) (Supplementary XXIII and VIII).
As a visual summary, Supplementary XXV displays the Bayesian predictive posterior densities for each week alongside the observed data for the four outbreaks during 2022–2023.
In addition, Figure \ref{fig: bayesianx=y} shows the Bayesian predictions plotted against the observed counts.

\section{Discussion}
In this study, we developed and evaluated a data parsimonious incidence versus cumulative cases (ICC) curves framework to forecast dengue outbreaks in Florida (2022–2023). 
Motivated by the limited generalizability and data requirements of many existing forecasting approaches, 
we extended the theoretical justification of the ICC approach beyond the basic SIR setting by proving that the same parabolic incidence–cumulative cases relation also holds for a complicated model of dengue transmission; the resulting ICC formulation remains a two-parameter model estimated from human incidence data.

We also used an established Bayesian framework based on the theory of the ICC method to quantify the uncertainty in reported human cases.
Our results show that both the Poisson model based on the deterministic ICC framework and its Bayesian extension reproduce key features of outbreak timing and trajectory with high predictive accuracy while requiring only the target year’s human case time series. 
Because the ICC model estimates only two parameters, it reduces estimation noise relative to higher-dimensional approaches. 
The Bayesian formulation, developed to capture the uncertainty arising from reporting and model fitting, yields further improvements in predictive performance.
The accurate results and findings (Section \ref{sec: results}) suggest that the parsimonious ICC-based model can mitigate some of the generalizability issues seen in location-specific forecasting frameworks. 
Because the only required input is the time series of new human cases in the target year, the approach does not rely on site-specific mosquito surveillance or detailed environmental covariates. In turn, it eliminates the need for mosquito-trap data and supports the use of a single model structure across different counties and outbreaks.
In addition, by reducing the parameter space to only two epidemiological parameters, the ICC model decreases the noise introduced by high-dimensional parameter estimation and simplifies model calibration. 
In our Florida application, both the censored Poisson and Bayesian models produced stable forecasts and good predictive accuracy, without requiring heavy model selection or complicated covariate choices.
Incorporating observation uncertainty directly into the Bayesian ICC model partially addresses the problem of noisy, under (over-) reported incidence.
In the Florida application, case reporting is highly accurate and we assumed a small observation variance.
As a result, the Bayesian posterior predictive domain intervals, which account for both process and reporting uncertainty, are well calibrated and relatively shorter than the domain of the distributions of censored Poisson model (Supplementary V-VIII).
Finally, because our method does not require a long outbreak history, it remains applicable to locations with limited historical data, which is crucial for newly affected regions or counties where dengue has only recently emerged.

\subsection{Ecology of dengue in Florida: interpret our results in that light}
Most locally acquired dengue cases in Florida have occurred in the southern part of the state, particularly in Miami-Dade and, to a lesser extent, Monroe County \cite{rey2014dengue,eisen2014temporal}.
To evaluate the ecological plausibility of our risk predictions in the Florida setting, we compared the model-based risk indices, the median of the censored Poisson model for each week prediction, with the weekly abundance of \textit{Aedes aegypti} in Miami-Dade County for the year 2022.
Although we only observe a single seasonal cycle and cannot assess whether these patterns are stable across years, the results are nevertheless consistent with field data and with other published studies.
Within the single-year time series, lagged Spearman correlations between weekly \textit{Aedes aegypti} abundance and model-predicted dengue risk (1 to 4-weeks ahead) reveal a consistent signal in the 8 to 10-week lag window ( Figure \ref{fig:forecast_2x2} and Supplementary XXIV).
An 8–10 week lag between vector dynamics and dengue risk is consistent with previous studies in Mexico, Sri Lanka, and southern Brazil \cite{liyanage2022assessing,eisen2014temporal,da2017meteorological}.
For 1--week ahead predictions, the correlation appears with lags of about 7 to 8 weeks (if $\text{p-value} \le 0.1$). 
For the 2-- and 3--week ahead predictions, similar patterns are observed with lags of about (7-9) weeks and (8-10), and for the 4--week ahead predictions the strongest correlations occur at lags of approximately 8 to 10 weeks.
Together, these results suggest that within this season, the seasonal pattern of mosquito abundance aligns strongly with the temporal pattern of predicted dengue risk at lags of roughly 8-10 weeks.
\vspace{0.5cm}
\subsection{Operational and public-health implications}

These lagged relationships and the accurate predictive performance of the model (quantified in Section \ref{sec: results}) may point to two possible implications for mitigation strategies.
First, the estimated 8-10 week delay between \textit{Aedes aegypti} abundance and dengue risk with the 1,2,3 and 4-weeks ahead implies an effective lead time on the order of approximately 2 months between elevated mosquito abundance and subsequent increases in human risk. 
In the Florida time series 2022-2023 (Figure \ref{fig:Florida 2022-23-seasons}), this suggests two main windows for intensified adulticiding for 2022: one early in the calendar year, preceding the initial increase in cases, and a second period beginning shortly after week 20, ahead of the larger late-season outbreak. 
Second, the correlation between mosquito abundance and the 4-week ahead predictions indicates that an operational threshold on the 4-week ahead forecast could be used to trigger targeted adulticiding, with approximately one month available to reduce adult mosquito abundance before the anticipated increase in transmission. 
Although these analyses highlight a clear correlation between mosquito abundance and predicted dengue risk, they are based on a single year of data. 
More precise estimates of the underlying ecological lag would require multi-year entomological surveillance and case data, which are currently limited for this setting.

\begin{figure}
    \centering
    \includegraphics[width=1\linewidth]{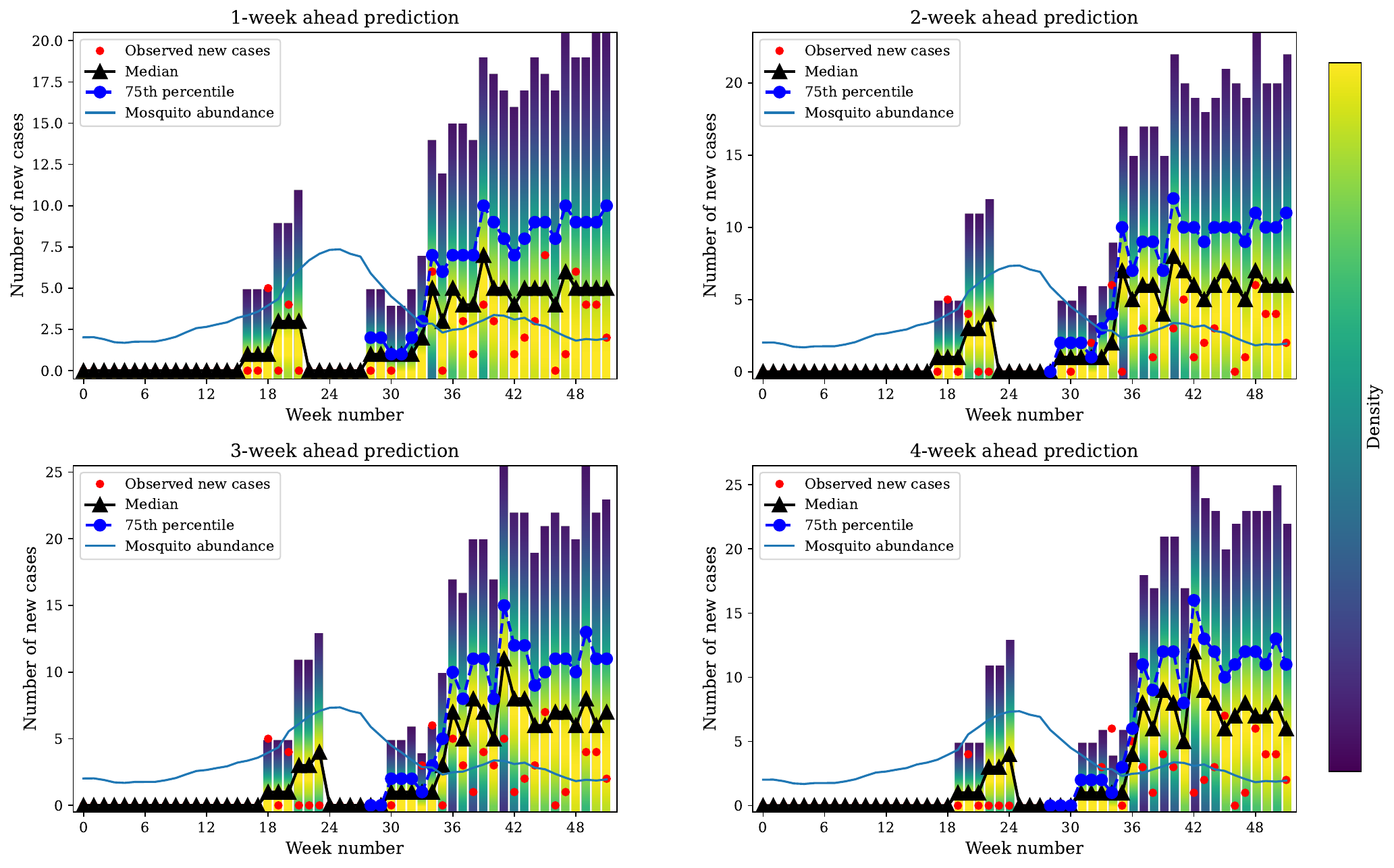}
    \caption{Each panel shows the censored Poisson model for the corresponding weekly prediction. The point prediction of risk for each week is shown by black triangles, representing the median of the censored Poisson distribution. Observed reported case counts are shown by red dots, and the blue curve depicts the smoothed weekly abundance of mosquitoes in traps. }
    \label{fig:forecast_2x2}
\end{figure}

Although detailed and time-resolved information on mosquito management in Miami-Dade is not available, the Florida Department of Health issued mosquito-borne dengue illness alerts in weeks 29–33 of 2022.
Probably, adulticiding was initiated or intensified during this advisory period and contributed to reducing both risk and incident cases.
Hence, the cases reported after this window can be interpreted as occurring under enhanced control. Under this assumption, the number of cases that might have occurred in the absence of such a program could plausibly lie closer to the upper quintile (for example, the 75th percentile) of the weekly risk distribution, as indicated by the blue circles in Figure \ref{fig:forecast_2x2}.

Beyond these implications for timing and control, the Bayesian framework also affects how we interpret uncertainty in forecasts.
Although the Bayesian model was primarily designed to account for reporting error rather than to replace the Poisson model, in most weeks its predictions are actually more precise: the posterior predictive intervals are shorter than the corresponding intervals from the censored Poisson model.
This is mainly because we assumed a very small variance for the observation error.
In other words, based on the testing procedures and protocols for patient diagnosis and protocols used by the CDC and Florida Department of Health explained in Section \ref{sec: observation model}, we specified a small reporting error, and this choice is directly reflected in the narrow posterior predictive intervals.
Accurate case reporting in this context refers to reliable identification and recording of clinically confirmed cases, and it does not imply that the reported counts equal the true number of infections in the population.

In addition, the Bayesian model provides evidence supporting the high precision of dengue case reporting by the Florida Department of Health.
We examined the posterior distribution of the true observations given the weekly reported counts, $\pi(O_p \mid \mathbf{x})$, and found that for each week, the posterior mean (median) was equal to or slightly greater than the corresponding reported case count, reflecting the small observation-error variance specified in Section \ref{sec: observation model} (for example, Supplementary IX).
The posterior that gives more density to the regions greater than the reported cases for each week is plausibly interpreted as capturing unobserved infections, such as asymptomatic cases or patients who did not seek care.
By contrast, a posterior with insisting on the regions falling below the reported counts would be difficult to reconcile with the high diagnostic specificity of Florida’s dengue surveillance, where false-positive case confirmations are expected to be rare.
Another noteworthy feature of the Bayesian framework appears near the end of outbreak 2: the Bayesian model extends the tail of the epidemic further in time than the censored Poisson model. This behavior of the weekly posterior is consistent with isolated weeks with nonzero reported cases after weeks 56, 60, and 65, and may indicate additional symptomatic infections that were not captured in other weeks or performing a mosquito control program (Supplementary XXV).

\subsection{Limitations and future work}
Despite these encouraging results, the proposed modeling framework has several limitations that warrant consideration. 
First, the model is entirely based on the reported incidence, which is known to underestimate true dengue infections due to mild and asymptomatic cases.
Although the Bayesian framework partially accommodates reporting uncertainty, we do not explicitly model under reporting mechanisms, and therefore our forecasts reflect only the dynamics of detected cases. 
Second, by design, this model operates only after human cases begin to be reported. This limitation can be addressed by using historical case time series to predict the onset of the outbreak before reporting begins.
Third, our evaluation is restricted to outbreaks in Florida between 2022 and 2023; future work should assess how well the ICC-based approach generalizes to other regions and to longer time horizons.\\

\section*{Acknowledgments}
This work has been supported by the USDA National Institute of Food and Agriculture, grant number 2022-67015-38059 through the NSF/NIH/USDA\\/BBSRC/BSF/NSFC Ecology and Evolution of Infectious Diseases Program, and the United States Department of Agriculture ARS under agreements number 58-3022-1-010 and 58-3022-3-025.

\section*{CRediT authorship contribution statement}
\noindent \textbf{Saman Hosseini}: Conceptualization, data curation, formal analysis, investigation, methodology, validation, visualization, writing original draft, writing review editing.\\
\noindent \textbf{Lee W. Cohnstaedt}: Conceptualization, methodology, resources, supervision, validation, writing original draft, writing, review editing.\\
\noindent \textbf{Caterina Scoglio}: Conceptualization, formal analysis, funding acquisition, methodology, project administration, supervision, validation, writing original draft, writing, review editing.
\nolinenumbers

\appendix
\section*{Appendix}
\section*{A-Four-serotype dengue model}\label{model}


\noindent All state variables are population counts (not fractions).
The total human and mosquito populations are assumed constant because the model does not include demographic processes such as births, natural deaths, or migration. 
The framework explicitly distinguishes primary and secondary infections, allowing individuals to progress from a first infection to a subsequent infection after recovery.
In addition, the model includes a separate compartment for mild and severe secondary infections, thereby accounting for the increased severity of the disease during reinfection.


\begingroup
\small
\setlength{\jot}{2pt}

\begin{equation*}
\lambda_{H,i}(t)=B_i\frac{V_i(t)}{N_M},
\qquad
\lambda_{M,i}(t)=A_i\frac{I_i(t)+Y_i(t)}{N_H},
\qquad i=1,2,3,4.
\end{equation*}

\begin{align*}
\dot S \;&=\; -S\sum_{i=1}^{4}\lambda_{H,i},\\
\dot I_i \;&=\; S\lambda_{H,i}-\gamma I_i,\qquad i=1,2,3,4, \\[2pt]
\dot R_1 \;&=\; \gamma I_1 - R_1\big(\sigma_2\lambda_{H,2}+\sigma_3\lambda_{H,3}+\sigma_4\lambda_{H,4}\big), \\
\dot R_2 \;&=\; \gamma I_2 - R_2\big(\sigma_1\lambda_{H,1}+\sigma_3\lambda_{H,3}+\sigma_4\lambda_{H,4}\big), \\
\dot R_3 \;&=\; \gamma I_3 - R_3\big(\sigma_1\lambda_{H,1}+\sigma_2\lambda_{H,2}+\sigma_4\lambda_{H,4}\big), \\
\dot R_4 \;&=\; \gamma I_4 - R_4\big(\sigma_1\lambda_{H,1}+\sigma_2\lambda_{H,2}+\sigma_3\lambda_{H,3}\big), \\[2pt]
\dot Y_1 \;&=\; (1-q)\sigma_1\lambda_{H,1}(R_2+R_3+R_4)-\gamma Y_1, \\
\dot Y_2 \;&=\; (1-q)\sigma_2\lambda_{H,2}(R_1+R_3+R_4)-\gamma Y_2, \\
\dot Y_3 \;&=\; (1-q)\sigma_3\lambda_{H,3}(R_1+R_2+R_4)-\gamma Y_3, \\
\dot Y_4 \;&=\; (1-q)\sigma_4\lambda_{H,4}(R_1+R_2+R_3)-\gamma Y_4, 
\end{align*}

\begin{align*}
\dot D \;=\; &\;
q\Big[
\sigma_1\lambda_{H,1}(R_2+R_3+R_4)
+\sigma_2\lambda_{H,2}(R_1+R_3+R_4)
\nonumber\\
&\qquad
+\sigma_3\lambda_{H,3}(R_1+R_2+R_4)
+\sigma_4\lambda_{H,4}(R_1+R_2+R_3)
\Big]
-\gamma D.
\label{eq:D}
\end{align*}

\begin{align*}
\dot R \;&=\; \gamma\Big(\sum_{i=1}^{4}Y_i + D\Big), \\[4pt]
\dot S_M \;&=\; -S_M\sum_{i=1}^{4}\lambda_{M,i}, \\
\dot V_i \;&=\; S_M\lambda_{M,i},\qquad i=1,2,3,4.
\end{align*}

\endgroup

\noindent \textbf{State variables:}

\vspace{0.5cm}
\noindent Humans:

\begin{itemize}
\item $S(t)$: never-infected (fully susceptible) humans.
\item $I_i(t)$: primary infectious humans with serotype $i$, $i=1,2,3,4$.
\item $R_i(t)$: humans recovered from primary infection with serotype $i$ (immune to $i$, susceptible to the other three).
\item $Y_i(t)$: secondary infectious humans currently infected with serotype $i$ (aggregated over any previous serotype $j\neq i$), mild class.
\item $D(t)$: severe secondary infectious humans.
\item $R(t)$: humans recovered after secondary infection.
\end{itemize}

Mosquitoes:
\begin{itemize}
\item $S_M(t)$: susceptible mosquitoes.
\item $V_i(t)$: mosquitoes infected with serotype $i$, $i=1,2,3,4$.
\end{itemize}

Totals (constants):
\[
N_H = S + \sum_{i=1}^4 I_i + \sum_{i=1}^4 R_i + \sum_{i=1}^4 Y_i + D + R,
\qquad
N_M = S_M + \sum_{i=1}^4 V_i.
\]

\paragraph{Parameters.}
\begin{itemize}
\item $B_i$: mosquito-to-human transmission coefficient for serotype $i$.
\item $A_i$: human-to-mosquito transmission coefficient for serotype $i$.
\item $\gamma$: human recovery rate (used for primary and secondary infection classes).
\item $\sigma_i$: susceptibility/enhancement multiplier for secondary infection with current serotype $i$.
\item $q\in[0,1]$: fraction of secondary infections routed to the severe class $D$ (the remaining $1-q$ go to the mild class $Y_i$).
\end{itemize}

\paragraph{Forces of infection (frequency-dependent).}
\[
\lambda_i^{H}(t) = B_i\frac{V_i(t)}{N_M}
\quad\text{(human infection hazard from infected mosquitoes of type $i$),}
\]
\[
\lambda_i^{M}(t) = A_i\frac{I_i(t)+Y_i(t)}{N_H}
\quad\text{(mosquito infection hazard from infectious humans of type $i$).}
\]

\paragraph{Summary of the modeling assumptions.}
\begin{itemize}
\item No demographic terms: $N_H$ and $N_M$ are constant.
\item Transmission is frequency-dependent, implemented by the ratios $V_i/N_M$ and $(I_i+Y_i)/N_H$.
\item Secondary infections are aggregated by the {current} serotype ($Y_i$), and severe outcomes are aggregated into $D$.
\end{itemize}

\section*{B-Incidence versus cumulative cases curves}\label{Appendix-B}

\noindent We consider the dengue transmission model in Section \ref{model}, which includes multiple serotypes and separate classes for primary and secondary infection. Our goal is to show that an ICC-type relation can be derived for this system. Specifically, our goal is to:

\begin{itemize}
\item defining a cumulative-case variable \(C(t)\);
\item defining the incidence as \(G(t)=\dot C(t)\);
\item showing that, early in the outbreak (near the disease-free equilibrium), incidence is approximately proportional to cumulative cases, i.e., \(G \propto C\);
\item then imposing the natural saturation condition that incidence vanishes at the final size \(L_0\), namely \(G(L_0)=0\), which leads to a simple logistic/parabolic ICC curve:
\[
G(C)\approx r\,C\left(1-\frac{C}{L_0}\right).
\]
\end{itemize}

\noindent These steps mirror the standard derivation for the basic SIR model. The key difference is that the dengue model has additional compartments, so the definitions of \(C(t)\) and the linearization must be handled carefully. 

\subsection*{Step 1. Choosing “cumulative cases” in the dengue model}
\vspace{0.5cm}
\noindent In the model, \(S(t)\) is the class of humans that have never been infected. The human population \(N_H\) is constant.
So, the most natural cumulative number of primary infections is:

\[
C(t)=N_H - S(t).
\]

\noindent Why this is the right definition (for “primary cases”):
\begin{itemize}
\item Every time someone leaves \(S\), that person becomes infected for the first time (goes into one of the \(I_i\) classes).
\item Therefore the decrease in \(S\) equals the accumulation of first infections.
\end{itemize}



\subsection*{Step 2. Exact incidence \(G(t)=\dot C(t)\)}
\vspace{0.5cm}
\noindent We define incidence as the rate of increase of cumulative cases:

\[
G(t)=\dot C(t).
\]

\noindent Since \(C=N_H-S\), and \(N_H\) is constant:

\[
\dot C(t)= -\dot S(t).
\]

\noindent So, to compute the incidence, we only need the equation for \(\dot S\).
From our model, the force of infection for serotype \(i\) is:

\[
\lambda_{H,i}(t)=B_i\frac{V_i(t)}{N_M}.
\]

\noindent Interpretation:
\begin{itemize}
\item \(V_i(t)\) = infectious mosquitoes carrying serotype \(i\)
\item \(N_M\) = total mosquito population (constant)
\item \(V_i/N_M\) = fraction of infectious mosquitoes
\item \(B_i\) = biting/transmission parameter (human infection per infectious mosquito fraction)
\end{itemize}

\noindent Our susceptible equation is (sum over all serotypes):

\[
\dot S
=
-S\big(\lambda_{H,1}+\lambda_{H,2}+\lambda_{H,3}+\lambda_{H,4}\big).
\]

\noindent Substitute \(\lambda_{H,i}=B_iV_i/N_M\):

\[
\dot S
=
-S\left(
\frac{B_1V_1}{N_M}+
\frac{B_2V_2}{N_M}+
\frac{B_3V_3}{N_M}+
\frac{B_4V_4}{N_M}
\right).
\]

\noindent Now, the incidence is:

\[
G(t)=\dot C(t)=-\dot S(t)
=
S(t)\left(
\frac{B_1V_1}{N_M}+
\frac{B_2V_2}{N_M}+
\frac{B_3V_3}{N_M}+
\frac{B_4V_4}{N_M}
\right).
\]

\noindent Or, more compactly:

\[
G(t)=S(t)\sum_{i=1}^{4}\frac{B_i}{N_M}\,V_i(t).
\]
\noindent This is an exact identity implied by the model; no approximation has been introduced at this stage. 
It shows that the incidence depends on the instantaneous system state, in particular on \(S(t)\) and the infected-mosquito compartments \(V_i(t)\). 
However, the right-hand side still involves multiple state variables, whereas an ICC formulation aims to express incidence as a function of cumulative cases \(C\) alone. 
To obtain such a reduced relation, we will derive approximations in a neighborhood of the relevant equilibrium.

\subsection*{Step 3. disease-free equilibrium (DFE)}
\vspace{0.5cm}
\noindent The disease-free equilibrium means no infection in humans and mosquitoes.
That means that all infected compartments are zero and everyone is susceptible.

\noindent Human compartments at DFE:
\[
S=N_H,\qquad I_i=0,\qquad R_i=0,\qquad Y_i=0,\qquad D=0,\qquad R=0.
\]

\noindent Mosquitoes compartments at DFE:
\[
S_M=N_M,\qquad V_i=0.
\]

\noindent This equilibrium is the starting point for linearization.

\subsection*{Step 4. Why secondary infection compartments do not affect the very early dynamics}

\noindent Our model has secondary infections \(Y_i\) and a severe class \(D\). Those equations have inflow terms that look like:

\begin{itemize}
\item “mosquito infection pressure” times “already immune humans,” for example:
\end{itemize}

\[
 \lambda_{H,i}(t)\times R_j(t).
\]

\noindent The key point: \(\lambda_{H,i}(t)\) is proportional to \(V_i(t)\), so these inflow terms contain products like:

\[
V_i(t)\,R_j(t).
\]

\noindent Now, near the DFE:
\begin{itemize}
\item \(V_i(t)\) is small (almost zero)
\item \(R_j(t)\) is also small (almost zero)
\end{itemize}

\noindent Hence the product \(V_iR_j\) is of second order (a product of small perturbations). Under a first-order linearization, only terms that are linear in the perturbations are retained, and all second-order products are neglected.
\noindent Consequently, in the initial phase of an outbreak:
\begin{itemize}
\item the inflows into secondary-infection compartments are negligible to first order; and
\item the early-time dynamics are dominated by the primary transmission loop linking \(I_i\) (infectious humans) and \(V_i\) (infectious mosquitoes).
\end{itemize}


\subsection*{Step 5. Linearization of the primary infection subsystem \((I_i,V_i)\)}

\noindent The primary human infection equation (for each serotype \(i\)) has the form:

\[
\dot I_i = \frac{B_i}{N_M} S V_i - \gamma I_i.
\]

\noindent Interpretation:
\begin{itemize}
\item new primary human infections are created by susceptible humans \(S\) infected by infected mosquitoes \(V_i\)
\item infected humans leave \(I_i\) at rate \(\gamma\)
\end{itemize}

\noindent The mosquito infection equation (for each serotype \(i\)) has the following form:

\[
\dot V_i = \frac{A_i}{N_H}(I_i+Y_i)S_M.
\]

\noindent Interpretation:
\begin{itemize}
\item mosquitoes become infected by biting infected humans
\item both primary infected humans \(I_i\) and secondary infected humans \(Y_i\) can infect mosquitoes (that is why \(I_i+Y_i\) appears)
\end{itemize}

\noindent Now linearize near DFE:

\begin{itemize}
\item \(S\approx N_H\)
\item \(S_M\approx N_M\)
\item \(Y_i\approx 0\) (from Step 3)
\end{itemize}

\noindent So, the linearized equations become:

\[
\dot I_i \approx \left(\frac{B_iN_H}{N_M}\right)V_i - \gamma I_i,
\]
\[
\dot V_i \approx \left(\frac{A_iN_M}{N_H}\right)I_i.
\]

\noindent This is a 2D linear system for each serotype \(i\).
In matrix form (clean, like the second picture you liked), we have the following relation:

\[
\frac{d}{dt}
\begin{bmatrix}
I_i\\
V_i
\end{bmatrix}
=
\begin{bmatrix}
-\gamma & \frac{B_iN_H}{N_M}\\
\frac{A_iN_M}{N_H} & 0
\end{bmatrix}
\begin{bmatrix}
I_i\\
V_i
\end{bmatrix}.
\]

\subsection*{Step 6. Early exponential growth rate \(r_i\)}

\noindent In linear systems, solutions behave as exponentials \(e^{rt}\) where \(r\) are eigenvalues.
So we solve for the eigenvalues \(r\) of the matrix:

\[
\begin{bmatrix}
-\gamma & \frac{B_iN_H}{N_M}\\
\frac{A_iN_M}{N_H} & 0
\end{bmatrix}.
\]

\noindent Eigenvalues satisfy:

\[
\det\!\left(
\begin{bmatrix}
-\gamma-r & \frac{B_iN_H}{N_M}\\
\frac{A_iN_M}{N_H} & -r
\end{bmatrix}
\right)=0.
\]

\noindent Compute determinant:

\[
(-\gamma-r)(-r)-
\left(\frac{B_iN_H}{N_M}\right)\left(\frac{A_iN_M}{N_H}\right)=0.
\]

\noindent Simplify the product:

\[
\left(\frac{B_iN_H}{N_M}\right)\left(\frac{A_iN_M}{N_H}\right)=A_iB_i.
\]

\noindent So we get:

\[
r^2+\gamma r - A_iB_i=0.
\]

\noindent Solve quadratic:

\[
r_i=\frac{-\gamma+\sqrt{\gamma^2+4A_iB_i}}{2}.
\]

\noindent This \(r_i\) is the early growth rate for serotype \(i\). If \(r_i>0\), the infection can invade (early exponential growth). If multiple serotypes are seeded, the one with the largest \(r_i\) dominates early growth.

\subsection*{Step 7. Conversion of the early exponential growth into an ICC linear law \(G\approx rC\)}

\noindent Recall:

\[
G(t)=\dot C(t)= S(t)\sum_{i=1}^4 \frac{B_i}{N_M}V_i(t).
\]

\noindent Near the start:
\[
S(t)\approx N_H,
\]
so
\[
G(t)\approx N_H\sum_{i=1}^4 \frac{B_i}{N_M}V_i(t).
\]

\noindent Now, from linearization (Step 5), each \(V_i(t)\) grows approximately like \(e^{r_it}\). If one mode dominates with rate \(r\), then the whole sum behaves as follows:

\[
G(t)\propto e^{rt}.
\]

\noindent But since \(G(t)=\dot C(t)\), if \(\dot C\propto e^{rt}\), then integration implies:

\[
C(t)\propto e^{rt}.
\]

\noindent Therefore, for early times (when the linearization is valid), we get:

\[
\dot C(t)\approx r\,C(t).
\]

\noindent This is the key ICC statement from linearization: on a plot of incidence vs cumulative cases, the curve must begin like a straight line with slope \(r\).

\subsection*{Step 8. Adding the saturation condition to obtain a parabolic/logistic ICC curve}

\noindent Let the final cumulative number of primary cases be:

\[
L_0=\lim_{t\to\infty}C(t)=N_H-S(\infty).
\]

\noindent At the end of the outbreak, the incidence goes to zero:

\[
G(L_0)=0.
\]

\noindent Now we have two constraints on the ICC function \(G(C)\):

\begin{enumerate}
\item Near \(C=0\): \(G(C)\approx rC\)
\item At \(C=L_0\): \(G(C)=0\)
\end{enumerate}

\noindent The simplest function satisfying both is:

\[
G(C)=\dot C \approx r\,C\left(1-\frac{C}{L_0}\right)
=\frac{r}{L_0}C(L_0-C).
\]

\noindent That is the inverted-parabola ICC relation, and it is exactly the logistic differential equation for cumulative cases \(C(t)\).

\subsection*{Step 8. Alternative case definition (including secondary infections).}
\noindent So far we have defined cumulative cases as first (primary) infections,
\[
C_{\mathrm{prim}}(t)=N_H-S(t), \qquad 
G_{\mathrm{prim}}(t)=\dot C_{\mathrm{prim}}(t)=-\dot S(t).
\]
\noindent If, instead, the reported case count is intended to include both primary and secondary infection
\emph{events}, we write the cumulative number of infection events as
\[
C_{\mathrm{all}}(t)=C_{\mathrm{prim}}(t)+C_{\mathrm{sec}}(t),
\qquad
G_{\mathrm{all}}(t)=\dot C_{\mathrm{all}}(t)=\dot C_{\mathrm{prim}}(t)+\dot C_{\mathrm{sec}}(t),
\]
\noindent where $\dot C_{\mathrm{sec}}(t)$ is defined as the total inflow into the secondary-infection classes
(the mild classes $Y_1,\dots,Y_4$ plus the severe class $D$). Concretely,
\[
\dot C_{\mathrm{sec}}(t)
=
\sum_{i=1}^{4}(1-q)\,\sigma_i\,\lambda_{H,i}(t)\sum_{j\neq i}R_j(t)
\;+\;
q\sum_{i=1}^{4}\sigma_i\,\lambda_{H,i}(t)\sum_{j\neq i}R_j(t).
\]

\noindent Near the disease-free equilibrium (DFE), $V_i(t)$ and $R_j(t)$ are small, and
$\lambda_{H,i}(t)$ is proportional to $V_i(t)$. 
Therefore, $\dot C_{\mathrm{sec}}(t)$ consists of products
of the form $V_i(t)R_j(t)$ and is second order under a first-order linearization, that is.
\[
\dot C_{\mathrm{sec}}(t)=o\!\left(\dot C_{\mathrm{prim}}(t)\right)
\qquad \text{as the outbreak begins.}
\]
\noindent Hence, the same early-time ICC linear law still holds for the alternative case definition.
\[
\dot C_{\mathrm{all}}(t)\approx r\,C_{\mathrm{all}}(t)
\qquad \text{(early outbreak / near DFE).}
\]

\noindent At later times, secondary infections can contribute non-negligibly and may deform the ICC relationship away
from an exact parabola, because $\dot C_{\mathrm{sec}}(t)$ depends on additional state variables
(e.g., the distribution of immunity across serotypes) and not on $C_{\mathrm{all}}(t)$ alone.
Nonetheless,
the early-growth condition $\dot C \approx rC$ as $C\to 0$ together with epidemic termination
$\dot C \to 0$ as $C\to L$ provide the same endpoint constraints that motivate a parabolic/logistic ICC
closure (i.e., $G(0)=0$, $G'(0)=r$, and $G(L)=0$ for $G(C)=\dot C$).

\section*{C-Metropolis--Hastings Algorithm for Sampling}

\noindent This section describes the Metropolis--Hastings (MH) procedure used to draw
samples from the posterior distribution
\(\pi(\mathbf{O}_{p}\mid T=t)\) (see Section~2.5) and to compute
posterior‑predictive values of \(C_{t}\).
\\
\\
\begin{algorithm}[H]
\small   
\SetKwInOut{Input}{Input}
\SetKwInOut{Output}{Output}

\Input{
\(\mathbf{x}_{p} =(x_{1},\dots,x_{p})^{\top}\) — reported counts\\
\(\delta_{e}\) — observation variance (prior scale for \(\mathbf O_p\))\\
\(L_{0}\) — prior mean of final size (from ICC fit)\\
\(\delta_{L}\) — prior variance of \(L\)\\
\(N\) (iterations), \(B\) (burn-in), \(s\) (thinning step)\\
\(\Sigma_{\text{prop}}^{O}\) — proposal covariance for \(\mathbf O_p\) (e.g.\ \(\sigma_{O}^{2}\mathbf I_{p}\))\\
\(\sigma_{\text{prop}}^{L}\) — proposal s.d.\ for \(L\)\\
ICC functions \(\mu(\mathbf O_p,L),\ \delta(\mathbf O_p,L)\)
}
\Output{Posterior draws \(\{\mathbf{O}^{(k)},\,L^{(k)},\,C_t^{(k)}\}\)}
\end{algorithm}

\begin{algorithm}[H]
\small   
\SetKwInOut{Input}{Input}
\SetKwInOut{Output}{Output}

\textbf{Initialisation:}\;
\(S \leftarrow \sum_{i=1}^{p} x_i\)  \tcp*{lower bound for \(L\)}
\(\mathbf{O}^{(0)}\leftarrow\mathbf{x}_{p}\) (all components $\ge 0$);\;
\(L^{(0)}\leftarrow \max\{L_{0},S\}\);\;
\(\ell^{(0)}\leftarrow\log\pi\!\bigl(\mathbf{O}^{(0)},L^{(0)}\mid T=t\bigr)\);\;
\(k_{\text{save}}\leftarrow 0\);

\For{$k=1$ \KwTo $N$}{
  \textbf{1~~Proposal}\;
  Draw $\bm{\varepsilon}_{O}\sim \mathcal{N}(\mathbf{0},\Sigma_{\text{prop}}^{O})$\;
  \(\mathbf{O}^{\ast} \gets \mathbf{O}^{(k-1)} + \bm{\varepsilon}_{O}\)\;
  \(\mathbf{O}^{\ast} \gets \max\!\bigl\{\mathbf{O}^{\ast},\,\mathbf 0\bigr\}\)
  \tcp*{reflect to keep $O_{i}^{\ast}\ge 0$}
  Draw $\varepsilon_{L}\sim \mathcal{N}(0,(\sigma_{\text{prop}}^{L})^{2})$\;
  \(L^{\ast} \gets L^{(k-1)} + \varepsilon_{L}\)\;

  \textbf{2~~Log-posterior ratio}\;
  \eIf{\(L^{\ast} < S\)}{
     \(\ell^{\ast}\leftarrow -\infty\)  \tcp*{outside support of truncated prior}
  }{
     \(\ell^{\ast}\leftarrow\log\pi\!\bigl(\mathbf{O}^{\ast},L^{\ast}\mid T=t\bigr)\);\;
  }
  \(\Delta\ell = \ell^{\ast}-\ell^{(k-1)}\);

  \textbf{3~~Accept / reject}\;
  Draw $u\sim\mathrm{Unif}(0,1)$\;
  \eIf{$\log u < \Delta\ell$}{
        \(\mathbf{O}^{(k)}\gets\mathbf{O}^{\ast}\);\;
        \(L^{(k)}\gets L^{\ast}\);\;
        \(\ell^{(k)}\gets\ell^{\ast}\);
  }{
        \(\mathbf{O}^{(k)}\gets\mathbf{O}^{(k-1)}\);\;
        \(L^{(k)}\gets L^{(k-1)}\);\;
        \(\ell^{(k)}\gets\ell^{(k-1)}\);
  }

  \textbf{4~~Store (after burn-in, with thinning)}\;
  \If{$k>B$ \textbf{and} $(k-B)\bmod s = 0$}{
      \(k_{\text{save}}\gets k_{\text{save}}+1\);\;
      \(\mathbf{O}^{(k_{\text{save}})}\gets\mathbf{O}^{(k)}\);\;
      \(L^{(k_{\text{save}})}\gets L^{(k)}\);\;
      \(C_t^{(k_{\text{save}})} \gets
        \dfrac{L^{(k)}}{1+\exp\!\bigl[-\delta\!\bigl(\mathbf{O}^{(k)},L^{(k)}\bigr)\,
              \bigl(t-\mu\!\bigl(\mathbf{O}^{(k)},L^{(k)}\bigr)\bigr)\bigr]}\).
  }
}
\Return{\(\bigl\{\mathbf{O}^{(k)},\,L^{(k)},\,C_t^{(k)}\bigr\}_{k=1}^{N_{\text{save}}}\)}

\caption{MH sampler for \(\pi(\mathbf O_p,L\mid T=t)\) (truncated–Gaussian prior for \(L\)) and draws of \(C_t\).}
\end{algorithm}

\section*{D-Logarithmic Score}
\noindent In this scheme, the predictive probability distribution for new cases is aggregated into epidemiologically motivated bins: the singleton $\{0\}$; the intervals $[1,5], [6,10], \dots, [46,50]$; the broader intervals $[50,100], [101,150], \dots, [201,250]$; and an out-of-range category. For a given forecast, let \(P\) denote the total predictive probability assigned to the bin that contains the observed count; the (natural) logarithmic score is
\[
s \;=\; \log P.
\]
Thus, a perfect forecast yields \(s=0\), and more negative values indicate worse performance.
If the score is less than -10, a default score of \(-10\) is assigned, and the prediction is classified as an outlier \cite{holcomb2023evaluation}.
For cumulative counts in week \(t\), we shifted the epidemiological bins by cumulative total throughout week \(t-1\), denoted \(R_{t-1}\).
Concretely, we added \(R_{t-1}\) to each bin end point because, for week \(t\), the support of the censored Poisson begins at \(R_{t-1}\). 
Thus, when the domain is \([R_{t-1}, \infty)\), the bins become the singleton \(\{R_{t-1}\}\); the intervals \([R_{t-1}+1,\,R_{t-1}+5], [R_{t-1}+6,\,R_{t-1}+10], \dots, [R_{t-1}+46,\,R_{t-1}+50]\); and the broader intervals \([R_{t-1}+50,\,R_{t-1}+100], [R_{t-1}+101,\,R_{t-1}+150], \dots, [R_{t-1}+201,\,R_{t-1}+250]\). As above, if the score is less than \(-10\), we truncate it to \(-10\) and label the forecast as an outlier.

\nocite{*}

\bibliography{bibl.bib}

@article{simmons2012dengue,
  title={Dengue},
  author={Simmons, Cameron P and Farrar, Jeremy J and van Vinh Chau, Nguyen and Wills, Bridget},
  journal={New England Journal of Medicine},
  volume={366},
  number={15},
  pages={1423--1432},
  year={2012},
  publisher={Mass Medical Soc}
}

@article{ross2010dengue,
  title={Dengue virus},
  author={Ross, Ted M},
  journal={Clinics in laboratory medicine},
  volume={30},
  number={1},
  pages={149},
  year={2010}
}

@article{schaffner2014dengue,
  title={Dengue and dengue vectors in the {W}{H}{O} {E}uropean region: past, present, and scenarios for the future},
  author={Schaffner, Francis and Mathis, Alexander},
  journal={The Lancet Infectious Diseases},
  volume={14},
  number={12},
  pages={1271--1280},
  year={2014},
  publisher={Elsevier}
}

@misc{WHO_dengue_factsheet,
  author = {World Health Organization},
  title  = {Dengue and severe dengue},
  year   = {2025},
  url    = {https://www.who.int/news-room/fact-sheets/detail/dengue-and-severe-dengue},
  note   = {Accessed 21 October 2025},
  publisher = {World Health Organization},
}

@article{paixao2015trends,
  title={Trends and factors associated with dengue mortality and fatality in {B}razil},
  author={Paix{\~a}o, Enny Santos and Costa, Maria da Concei{\c{c}}{\~a}o Nascimento and Rodrigues, Laura Cunha and Rasella, Davide and Cardim, Luciana Lobato and Brasileiro, Alcione Cunha and Teixeira, Maria Gloria Lima Cruz},
  journal={Revista da Sociedade Brasileira de Medicina Tropical},
  volume={48},
  number={4},
  pages={399--405},
  year={2015},
  publisher={SciELO Brasil}
}

@article{lai2017pharmacological,
  title={Pharmacological intervention for dengue virus infection},
  author={Lai, Jenn-Haung and Lin, Yi-Ling and Hsieh, Shie-Liang},
  journal={Biochemical Pharmacology},
  volume={129},
  pages={14--25},
  year={2017},
  publisher={Elsevier}
}

@article{thomas2023new,
  title={Is new dengue vaccine efficacy data a relief or cause for concern?},
  author={Thomas, Stephen J},
  journal={npj Vaccines},
  volume={8},
  number={1},
  pages={55},
  year={2023},
  publisher={Nature Publishing Group UK London}
}

@article{rather2017prevention,
  title={Prevention and control strategies to counter dengue virus infection},
  author={Rather, Irfan A and Parray, Hilal A and Lone, Jameel B and Paek, Woon K and Lim, Jeongheui and Bajpai, Vivek K and Park, Yong-Ha},
  journal={Frontiers in cellular and infection microbiology},
  volume={7},
  pages={336},
  year={2017},
  publisher={Frontiers Media SA}
}

@article{knerer2015impact,
  title={Impact of combined vector-control and vaccination strategies on transmission dynamics of dengue fever: a model-based analysis},
  author={Knerer, Gerhart and Currie, Christine SM and Brailsford, Sally C},
  journal={Health care management science},
  volume={18},
  number={2},
  pages={205--217},
  year={2015},
  publisher={Springer}
}

@article{ranathunge2022development,
  title={Development of the {S}terile {I}nsect {T}echnique to control the dengue vector {A}edes aegypti ({L}innaeus) in {S}ri {L}anka},
  author={Ranathunge, Tharaka and Harishchandra, Jeevanie and Maiga, Hamidou and Bouyer, Jeremy and Gunawardena, YI Nilmini Silva and Hapugoda, Menaka},
  journal={Plos one},
  volume={17},
  number={4},
  pages={e0265244},
  year={2022},
  publisher={Public Library of Science San Francisco, CA USA}
}

@article{bowman2014assessing,
  title={Assessing the relationship between vector indices and dengue transmission: a systematic review of the evidence},
  author={Bowman, Leigh R and Runge-Ranzinger, Silvia and McCall, Philip J},
  journal={PLoS neglected tropical diseases},
  volume={8},
  number={5},
  pages={e2848},
  year={2014},
  publisher={Public Library of Science San Francisco, USA}
}

@article{sasmita2021ovitrap,
  title={Ovitrap surveillance of dengue vector mosquitoes in bandung city, west java province, {I}ndonesia},
  author={Sasmita, Hadian Iman and Neoh, Kok-Boon and Yusmalinar, Sri and Anggraeni, Tjandra and Chang, Niann-Tai and Bong, Lee-Jin and Putra, Ramadhani Eka and Sebayang, Amelia and Silalahi, Christina Natalina and Ahmad, Intan and others},
  journal={PLoS neglected tropical diseases},
  volume={15},
  number={10},
  pages={e0009896},
  year={2021},
  publisher={Public Library of Science San Francisco, CA USA}
}

@article{ong2021adult,
  title={Adult {A}edes abundance and risk of dengue transmission},
  author={Ong, Janet and Aik, Joel and Ng, Lee Ching},
  journal={PLoS Neglected Tropical Diseases},
  volume={15},
  number={6},
  pages={e0009475},
  year={2021},
  publisher={Public Library of Science San Francisco, CA USA}
}

@article{lega2017aedes,
  title={Aedes aegypti ({D}iptera: {C}ulicidae) abundance model improved with relative humidity and precipitation-driven egg hatching},
  author={Lega, Joceline and Brown, Heidi E and Barrera, Roberto},
  journal={Journal of medical entomology},
  volume={54},
  number={5},
  pages={1375--1384},
  year={2017},
  publisher={Oxford University Press}
}

@article{tran2013rainfall,
  title={A rainfall-and temperature-driven abundance model for {A}edes albopictus populations},
  author={Tran, Annelise and L’ambert, Gr{\'e}gory and Lacour, Guillaume and Beno{\^\i}t, Romain and Demarchi, Marie and Cros, Myriam and Cailly, Priscilla and Aubry-Kientz, M{\'e}laine and Balenghien, Thomas and Ezanno, Pauline},
  journal={International journal of environmental research and public health},
  volume={10},
  number={5},
  pages={1698--1719},
  year={2013},
  publisher={MDPI}
}

@article{serpa2013study,
  title={Study of the distribution and abundance of the eggs of {A}edes aegypti and {A}edes albopictus according to the habitat and meteorological variables, municipality of {S}{\~a}o {S}ebasti{\~a}o, {S}{\~a}o {P}aulo {S}tate, {B}razil},
  author={Serpa, L{\'\i}gia Leandro Nunes and Monteiro Marques, Gisela Rita Alvarenga and de Lima, Ana Paula and Voltolini, J{\'u}lio Cesar and Arduino, Marylene de Brito and Barbosa, Gerson Laurindo and Andrade, Valmir Roberto and de Lima, Virg{\'\i}lia Luna Castor},
  journal={Parasites \& Vectors},
  volume={6},
  number={1},
  pages={321},
  year={2013},
  publisher={Springer}
}

@article{wang2023spatial,
  title={Spatial and temporal analyses of the influences of meteorological and environmental factors on {A}edes albopictus ({D}iptera: {C}ulicidae) population dynamics during the peak abundance period at a city scale},
  author={Wang, Fei and Zhu, Yiyi and Zhang, Hengduan and Fan, Junhua and Leng, Peien and Zhou, Ji and Yao, Shenjun and Yang, Dandan and Liu, Yao and Wang, Jingjing and others},
  journal={Acta Tropica},
  volume={245},
  pages={106964},
  year={2023},
  publisher={Elsevier}
}

@article{hii2012forecast,
  title={Forecast of dengue incidence using temperature and rainfall},
  author={Hii, Yien Ling and Zhu, Huaiping and Ng, Nawi and Ng, Lee Ching and Rockl{\"o}v, Joacim},
  journal={PLoS neglected tropical diseases},
  volume={6},
  number={11},
  pages={e1908},
  year={2012},
  publisher={Public Library of Science San Francisco, USA}
}

@article{descloux2012climate,
  title={Climate-based models for understanding and forecasting dengue epidemics},
  author={Descloux, Elodie and Mangeas, Morgan and Menkes, Christophe Eug{\`e}ne and Lengaigne, Matthieu and Leroy, Anne and Tehei, Temaui and Guillaumot, Laurent and Teurlai, Magali and Gourinat, Ann-Claire and Benzler, Justus and others},
  journal={PLoS neglected tropical diseases},
  volume={6},
  number={2},
  pages={e1470},
  year={2012},
  publisher={Public Library of Science San Francisco, USA}
}

@article{lowe2014dengue,
  title={Dengue outlook for the {W}orld {C}up in {B}razil: an early warning model framework driven by real-time seasonal climate forecasts},
  author={Lowe, Rachel and Barcellos, Christovam and Coelho, Caio AS and Bailey, Trevor C and Coelho, Giovanini Evelim and Graham, Richard and Jupp, Tim and Ramalho, Walter Massa and Carvalho, Marilia S{\'a} and Stephenson, David B and others},
  journal={The Lancet infectious diseases},
  volume={14},
  number={7},
  pages={619--626},
  year={2014},
  publisher={Elsevier}
}

@article{shi2016three,
  title={Three-month real-time dengue forecast models: an early warning system for outbreak alerts and policy decision support in {S}ingapore},
  author={Shi, Yuan and Liu, Xu and Kok, Suet-Yheng and Rajarethinam, Jayanthi and Liang, Shaohong and Yap, Grace and Chong, Chee-Seng and Lee, Kim-Sung and Tan, Sharon SY and Chin, Christopher Kuan Yew and others},
  journal={Environmental health perspectives},
  volume={124},
  number={9},
  pages={1369--1375},
  year={2016},
  publisher={National Institute of Environmental Health Sciences}
}

@article{ranstam2018lasso,
  title={{L}{A}{S}{S}{O} regression},
  author={Ranstam, Jonas and Cook, Jonathan A},
  journal={Journal of British Surgery},
  volume={105},
  number={10},
  pages={1348--1348},
  year={2018},
  publisher={Oxford University Press}
}

@article{lowe2016evaluating,
  title={Evaluating probabilistic dengue risk forecasts from a prototype early warning system for {B}razil},
  author={Lowe, Rachel and Coelho, Caio AS and Barcellos, Christovam and Carvalho, Marilia Sa and Catao, Rafael De Castro and Coelho, Giovanini E and Ramalho, Walter Massa and Bailey, Trevor C and Stephenson, David B and Rodo, Xavier},
  journal={Elife},
  volume={5},
  pages={e11285},
  year={2016},
  publisher={eLife Sciences Publications, Ltd}
}

@article{hussain2018early,
  title={Early warning and response system ({E}{W}{A}{R}{S}) for dengue outbreaks: {R}ecent advancements towards widespread applications in critical settings},
  author={Hussain-Alkhateeb, Laith and Kroeger, Axel and Olliaro, Piero and Rockl{\"o}v, Joacim and Sewe, Maquins Odhiambo and Tejeda, Gustavo and Benitez, David and Gill, Balvinder and Hakim, S Lokman and Gomes Carvalho, Roberta and others},
  journal={PloS one},
  volume={13},
  number={5},
  pages={e0196811},
  year={2018},
  publisher={Public Library of Science San Francisco, CA USA}
}

@article{aburas2010dengue,
  title={Dengue confirmed-cases prediction: A neural network model},
  author={Aburas, Hani M and Cetiner, B Gultekin and Sari, Murat},
  journal={Expert Systems with Applications},
  volume={37},
  number={6},
  pages={4256--4260},
  year={2010},
  publisher={Elsevier}
}

@article{adde2016predicting,
  title={Predicting dengue fever outbreaks in French Guiana using climate indicators},
  author={Adde, Antoine and Roucou, Pascal and Mangeas, Morgan and Ardillon, Vanessa and Desenclos, Jean-Claude and Rousset, Dominique and Girod, Romain and Briolant, S{\'e}bastien and Quenel, Philippe and Flamand, Claude},
  journal={PLoS neglected tropical diseases},
  volume={10},
  number={4},
  pages={e0004681},
  year={2016},
  publisher={Public Library of Science San Francisco, CA USA}
}

@article{gultekin2010neural,
  title={Development of data-driven machine learning models and their potential role in predicting dengue outbreak},
  author={Mazhar, Bushra and Ali, Nazish Mazhar and Manzoor, Farkhanda and Khan, Muhammad Kamran and Nasir, Muhammad and Ramzan, Muhammad},
  journal={Journal of Vector Borne Diseases},
  volume={61},
  number={4},
  pages={503--514},
  year={2024},
  publisher={Medknow}
}

@article{edussuriya2021accurate,
  title={An accurate mathematical model predicting number of dengue cases in tropics},
  author={Edussuriya, Chathurangi and Deegalla, Sampath and Gawarammana, Indika},
  journal={PLoS neglected tropical diseases},
  volume={15},
  number={11},
  pages={e0009756},
  year={2021},
  publisher={Public Library of Science San Francisco, CA USA}
}

@article{buczak2018ensemble,
  title={Ensemble method for dengue prediction},
  author={Buczak, Anna L and Baugher, Benjamin and Moniz, Linda J and Bagley, Thomas and Babin, Steven M and Guven, Erhan},
  journal={PloS one},
  volume={13},
  number={1},
  pages={e0189988},
  year={2018},
  publisher={Public Library of Science San Francisco, CA USA}
}

@article{siriyasatien2016analysis,
  title={Analysis of significant factors for dengue fever incidence prediction},
  author={Siriyasatien, Padet and Phumee, Atchara and Ongruk, Phatsavee and Jampachaisri, Katechan and Kesorn, Kraisak},
  journal={BMC bioinformatics},
  volume={17},
  number={1},
  pages={166},
  year={2016},
  publisher={Springer}
}

@article{salim2021prediction,
  title={Prediction of dengue outbreak in Selangor Malaysia using machine learning techniques},
  author={Salim, Nurul Azam Mohd and Wah, Yap Bee and Reeves, Caitlynn and Smith, Madison and Yaacob, Wan Fairos Wan and Mudin, Rose Nani and Dapari, Rahmat and Sapri, Nik Nur Fatin Fatihah and Haque, Ubydul},
  journal={Scientific reports},
  volume={11},
  number={1},
  pages={939},
  year={2021},
  publisher={Nature Publishing Group UK London}
}

@article{deb2017ensemble,
  title={An ensemble prediction approach to weekly Dengue cases forecasting based on climatic and terrain conditions},
  author={Deb, Sougata and Acebedo, Cleta Milagros Libre and Dhanapal, Gomathypriya and Heng, Chua Matthew Chin},
  journal={Journal of Health and Social Sciences},
  volume={2},
  number={3},
  pages={257--272},
  year={2017}
}

@article{majeed2023prediction,
  title={Prediction of dengue cases using the attention-based long short-term memory (LSTM) approach},
  author={Majeed, Mokhalad A and Shafri, Helmi ZM and Wayayok, Aimrun and Zulkafli, Zed and others},
  journal={Geospatial health},
  volume={18},
  number={1},
  year={2023}
}

@article{ramadona2016prediction,
  title={Prediction of dengue outbreaks based on disease surveillance and meteorological data},
  author={Ramadona, Aditya Lia and Lazuardi, Lutfan and Hii, Yien Ling and Holmner, {\AA}sa and Kusnanto, Hari and Rockl{\"o}v, Joacim},
  journal={PloS one},
  volume={11},
  number={3},
  pages={e0152688},
  year={2016},
  publisher={Public Library of Science San Francisco, CA USA}
}

@article{althouse2011prediction,
  title={Prediction of dengue incidence using search query surveillance},
  author={Althouse, Benjamin M and Ng, Yih Yng and Cummings, Derek AT},
  journal={PLoS neglected tropical diseases},
  volume={5},
  number={8},
  pages={e1258},
  year={2011},
  publisher={Public Library of Science San Francisco, USA}
}

@article{bal2020modeling,
  title={Modeling and prediction of dengue occurrences in Kolkata, India, based on climate factors},
  author={Bal, Sourabh and Sodoudi, Sahar},
  journal={International journal of biometeorology},
  volume={64},
  number={8},
  pages={1379--1391},
  year={2020},
  publisher={Springer}
}

@article{chen2018neighbourhood,
  title={Neighbourhood level real-time forecasting of dengue cases in tropical urban Singapore},
  author={Chen, Yirong and Ong, Janet Hui Yi and Rajarethinam, Jayanthi and Yap, Grace and Ng, Lee Ching and Cook, Alex R},
  journal={BMC medicine},
  volume={16},
  number={1},
  pages={129},
  year={2018},
  publisher={Springer}
}

@article{siriyasatien2018dengue,
  title={Dengue epidemics prediction: A survey of the state-of-the-art based on data science processes},
  author={Siriyasatien, Padet and Chadsuthi, Sudarat and Jampachaisri, Katechan and Kesorn, Kraisak},
  journal={IEEE Access},
  volume={6},
  pages={53757--53795},
  year={2018},
  publisher={IEEE}
}

@article{buczak2012data,
  title={A data-driven epidemiological prediction method for dengue outbreaks using local and remote sensing data},
  author={Buczak, Anna L and Koshute, Phillip T and Babin, Steven M and Feighner, Brian H and Lewis, Sheryl H},
  journal={BMC medical informatics and decision making},
  volume={12},
  number={1},
  pages={124},
  year={2012},
  publisher={Springer}
}

@article{focks1997pupal,
  title={Pupal survey: an epidemiologically significant surveillance method for Aedes aegypti: an example using data from Trinidad.},
  author={Focks, Dana A and Chadee, Dave D},
  journal={The American journal of tropical medicine and hygiene},
  volume={56},
  number={2},
  pages={159--167},
  year={1997}
}

@article{seng2009pupal,
  title={Pupal sampling for Aedes aegypti (L.) surveillance and potential stratification of dengue high-risk areas in Cambodia},
  author={Seng, Chang M and Setha, To and Nealon, Joshua and Socheat, Duong},
  journal={Tropical Medicine \& International Health},
  volume={14},
  number={10},
  pages={1233--1240},
  year={2009},
  publisher={Wiley Online Library}
}

@article{pham2011ecological,
  title={Ecological factors associated with dengue fever in a Central Highlands province, Vietnam},
  author={Pham, Hau V and Doan, Huong TM and Phan, Thao TT and Tran Minh, Nguyen N},
  journal={BMC infectious diseases},
  volume={11},
  number={1},
  pages={172},
  year={2011},
  publisher={Springer}
}

@article{naish2014climate,
  title={Climate change and dengue: a critical and systematic review of quantitative modelling approaches},
  author={Naish, Suchithra and Dale, Pat and Mackenzie, John S and McBride, John and Mengersen, Kerrie and Tong, Shilu},
  journal={BMC infectious diseases},
  volume={14},
  number={1},
  pages={167},
  year={2014},
  publisher={Springer}
}

@article{gharbi2011time,
  title={Time series analysis of dengue incidence in Guadeloupe, French West Indies: forecasting models using climate variables as predictors},
  author={Gharbi, Myriam and Quenel, Philippe and Gustave, Jo{\"e}l and Cassadou, Sylvie and Ruche, Guy La and Girdary, Laurent and Marrama, Laurence},
  journal={BMC infectious diseases},
  volume={11},
  number={1},
  pages={166},
  year={2011},
  publisher={Springer}
}

@article{mclennan2014complex,
  title={Complex behaviour in a dengue model with a seasonally varying vector population},
  author={McLennan-Smith, Timothy A and Mercer, Geoffry N},
  journal={Mathematical biosciences},
  volume={248},
  pages={22--30},
  year={2014},
  publisher={Elsevier}
}

@article{karim2012climatic,
  title={Climatic factors influencing dengue cases in Dhaka city: a model for dengue prediction},
  author={Karim, Md Nazmul and Munshi, Saif Ullah and Anwar, Nazneen and Alam, Md Shah},
  journal={Indian journal of medical research},
  volume={136},
  number={1},
  pages={32--39},
  year={2012},
  publisher={Medknow}
}

@article{wu2009higher,
  title={Higher temperature and urbanization affect the spatial patterns of dengue fever transmission in subtropical Taiwan},
  author={Wu, Pei-Chih and Lay, Jinn-Guey and Guo, How-Ran and Lin, Chuan-Yao and Lung, Shih-Chun and Su, Huey-Jen},
  journal={Science of the total Environment},
  volume={407},
  number={7},
  pages={2224--2233},
  year={2009},
  publisher={Elsevier}
}

@article{reiter2003texas,
  title={Texas lifestyle limits transmission of dengue virus},
  author={Reiter, Paul and Lathrop, Sarah and Bunning, Michel and Biggerstaff, Brad and Singer, Daniel and Tiwari, Tejpratap and Baber, Laura and Amador, Manuel and Thirion, Jaime and Hayes, Jack and others},
  journal={Emerging infectious diseases},
  volume={9},
  number={1},
  pages={86},
  year={2003}
}

@article{wongkoon2013weather,
  title={Weather factors influencing the occurrence of dengue fever in Nakhon Si Thammarat, Thailand.},
  author={Wongkoon, S and Jaroensutasinee, M and Jaroensutasinee, K},
  journal={Tropical Biomedicine},
  year={2013}
}

@article{xu2014statistical,
  title={Statistical modeling reveals the effect of absolute humidity on dengue in Singapore},
  author={Xu, Hai-Yan and Fu, Xiuju and Lee, Lionel Kim Hock and Ma, Stefan and Goh, Kee Tai and Wong, Jiancheng and Habibullah, Mohamed Salahuddin and Lee, Gary Kee Khoon and Lim, Tian Kuay and Tambyah, Paul Anantharajah and others},
  journal={PLoS neglected tropical diseases},
  volume={8},
  number={5},
  pages={e2805},
  year={2014},
  publisher={Public Library of Science San Francisco, USA}
}

@article{machado2012empirical,
  title={Empirical mapping of suitability to dengue fever in Mexico using species distribution modeling},
  author={Machado-Machado, Elia Axinia},
  journal={Applied Geography},
  volume={33},
  pages={82--93},
  year={2012},
  publisher={Elsevier}
}

@article{runge2008does,
  title={What does dengue disease surveillance contribute to predicting and detecting outbreaks and describing trends?},
  author={Runge-Ranzinger, Silvia and Horstick, Olaf and Marx, Michael and Kroeger, Axel},
  journal={Tropical Medicine \& International Health},
  volume={13},
  number={8},
  pages={1022--1041},
  year={2008},
  publisher={Wiley Online Library}
}

@inproceedings{aburas2007aburas,
  title={ABURAS index: A statistically developed index for dengue-transmitting vector population prediction},
  author={Aburas, Hani M},
  booktitle={Proceedings of world academy of science, engineering and technology},
  volume={23},
  pages={151--154},
  year={2007}
}

@article{strickman2003dengue,
  title={Dengue and its vectors in Thailand: calculated transmission risk from total pupal counts of Aedes aegypti and association of wing-length measurements with aspects of the larval habitat.},
  author={Strickman, Daniel and Kittayapong, Pattamaporn},
  journal={The American Journal of Tropical Medicine and Hygiene},
  volume={68},
  number={2},
  pages={209--217},
  year={2003}
}

@article{thongrungkiat2012natural,
  title={Natural transovarial dengue virus infection rate in both sexes of dark and pale forms of Aedes aegypti from an urban area of Bangkok, Thailand},
  author={Thongrungkiat, Supatra and Wasinpiyamongkol, Ladawan and Maneekan, Pannamas and Prummongkol, Samrerng and Samung, Yudthana},
  journal={Southeast Asian Journal of Tropical Medicine \& Public Health},
  volume={43},
  number={5},
  pages={1146--1152},
  year={2012}
}

@article{ponlawat2007age,
  title={Age and body size influence male sperm capacity of the dengue vector Aedes aegypti (Diptera: Culicidae)},
  author={Ponlawat, Alongkot and Harrington, Laura C},
  journal={Journal of medical entomology},
  volume={44},
  number={3},
  pages={422--426},
  year={2007},
  publisher={Oxford University Press Oxford, UK}
}

@article{limkittikul2014epidemiological,
  title={Epidemiological trends of dengue disease in Thailand (2000--2011): a systematic literature review},
  author={Limkittikul, Kriengsak and Brett, Jeremy and L'Azou, Ma{\"\i}na},
  journal={PLoS neglected tropical diseases},
  volume={8},
  number={11},
  pages={e3241},
  year={2014},
  publisher={Public Library of Science San Francisco, USA}
}

@article{chompoosri2012seasonal,
  title={Seasonal monitoring of dengue infection in Aedes aegypti and serological feature of patients with suspected dengue in 4 central provinces of Thailand},
  author={Chompoosri, Jakkrawarn and Thavara, Usavadee and Tawatsin, Apiwat and Anantapreecha, Surapee and Siriyasatien, Padet},
  journal={The Thai Journal of Veterinary Medicine},
  volume={42},
  number={2},
  pages={185--193},
  year={2012},
  publisher={Faculty of Veterinary Science, Chulalongkorn University}
}

@article{thavara2015biology,
  title={Biology of dengue vectors and serotypes of dengue virus in infectious cycle in Thailand},
  author={Thavara, U and Bhakdeenuan, P and Thawatsin, A and Chompoosri, J and Khumsawads, C and Phusup, Y and Phumee, A and Pengsakul, T and Siriyasatien, P and Sangkitporn, S},
  journal={Bull. Dept. Med. Sci},
  volume={57},
  number={2},
  pages={186--196},
  year={2015}
}

@article{cdc2007dengue,
  title={Dengue hemorrhagic fever--US-Mexico border, 2005},
  author={CDC},
  journal={MMWR Morb. Mortal. Wkly. Rep.},
  volume={56},
  pages={785--789},
  year={2007}
}

@article{knowlton2009fever,
  title={Fever pitch: Mosquito-borne dengue fever threat spreading in the Americas},
  author={Knowlton, Kim and Solomon, Gina and Rotkin-Ellman, Miriam and Council, NRD},
  journal={New York: Natural Resources Defense Council},
  year={2009}
}

@article{rund2019rescuing,
  title={Rescuing troves of hidden ecological data to tackle emerging mosquito-borne diseases},
  author={Rund, Samuel SC and Moise, Imelda K and Beier, John C and Martinez, Micaela Elvira},
  journal={Journal of the American Mosquito Control Association},
  volume={35},
  number={1},
  pages={75--83},
  year={2019},
  publisher={American Mosquito Control Association, Inc.}
}

@article{biswas2014dengue,
  title={Dengue fever in a rural area of West Bengal, India, 2012: an outbreak investigation},
  author={Biswas, Dilip K and Bhunia, Rama and Basu, Mausumi},
  journal={WHO South-East Asia Journal of Public Health},
  volume={3},
  number={1},
  pages={46--50},
  year={2014},
  publisher={Medknow}
}

@article{c2015surveillance,
  title={Surveillance of dengue vectors using spatio-temporal Bayesian modeling},
  author={C. Costa, Ana Carolina and Code{\c{c}}o, Cl{\'a}udia T and Hon{\'o}rio, Nildimar A and Pereira, Gl{\'a}ucio R and N. Pinheiro, Carmen F{\'a}tima and Nobre, Aline A},
  journal={BMC medical informatics and decision making},
  volume={15},
  number={1},
  pages={93},
  year={2015},
  publisher={Springer}
}

@article{thiruchelvam2018correlation,
  title={Correlation analysis of air pollutant index levels and dengue cases across five different zones in Selangor, Malaysia},
  author={Thiruchelvam, Loshini and Dass, Sarat C and Zaki, Rafdzah and Yahya, Abqariyah and Asirvadam, Vijanth S and others},
  journal={Geospatial health},
  volume={13},
  number={1},
  year={2018}
}

@article{beatty2005travel,
  title={Travel-Associated Dengue Infections--United States, 2001-2004.},
  author={Beatty, ME and Vorndam, V and Hunsperger, EA and Mu{\~n}oz, JL and Clark, GG},
  journal={MMWR: Morbidity \& Mortality Weekly Report},
  volume={54},
  number={22},
  year={2005}
}

@article{baaten2011travel,
  title={Travel-related dengue virus infection, the Netherlands, 2006--2007},
  author={Baaten, Gijs GG and Sonder, Gerard JB and Zaaijer, Hans L and van Gool, Tom and Kint, Joan APCM and van den Hoek, Anneke},
  journal={Emerging infectious diseases},
  volume={17},
  number={5},
  pages={821},
  year={2011}
}

@article{ratnam2013dengue,
  title={Dengue fever and international travel},
  author={Ratnam, Irani and Leder, Karin and Black, Jim and Torresi, Joseph},
  journal={Journal of travel medicine},
  volume={20},
  number={6},
  pages={384--393},
  year={2013},
  publisher={Oxford University Press Oxford, UK}
}

@article{hanna2006multiple,
  title={Multiple outbreaks of dengue serotype 2 in north Queensland, 2003/04},
  author={Hanna, Jeffrey N and Ritchie, Scott A and Richards, Ann R and Taylor, Carmel T and Pyke, Alyssa T and Montgomery, Brian L and Piispanen, John P and Morgan, Anna K and Humphreys, Jan L},
  journal={Australian and New Zealand journal of public health},
  volume={30},
  number={3},
  pages={220--225},
  year={2006},
  publisher={Wiley Online Library}
}

@article{teichmann2004dengue,
  title={Dengue virus infection in travellers returning to Berlin, Germany: clinical, laboratory, and diagnostic aspects},
  author={Teichmann, Dieter and G{\"o}bels, Klaus and Niedrig, Matthias and Grobusch, Martin P},
  journal={Acta Tropica},
  volume={90},
  number={1},
  pages={87--95},
  year={2004},
  publisher={Elsevier}
}

@article{reiner2014socially,
  title={Socially structured human movement shapes dengue transmission despite the diffusive effect of mosquito dispersal},
  author={Reiner Jr, Robert C and Stoddard, Steven T and Scott, Thomas W},
  journal={Epidemics},
  volume={6},
  pages={30--36},
  year={2014},
  publisher={Elsevier}
}

@article{gardner2012predictive,
  title={A predictive spatial model to quantify the risk of air-travel-associated dengue importation into the United States and Europe},
  author={Gardner, Lauren M and Fajardo, David and Waller, S Travis and Wang, Ophelia and Sarkar, Sahotra},
  journal={Journal of tropical medicine},
  volume={2012},
  number={1},
  pages={103679},
  year={2012},
  publisher={Wiley Online Library}
}

@article{wilke2019community,
  title={Community composition and year-round abundance of vector species of mosquitoes make Miami-Dade County, Florida a receptive gateway for arbovirus entry to the United States},
  author={Wilke, Andr{\'e} BB and Vasquez, Chalmers and Medina, Johana and Carvajal, Augusto and Petrie, William and Beier, John C},
  journal={Scientific reports},
  volume={9},
  number={1},
  pages={8732},
  year={2019},
  publisher={Nature Publishing Group UK London}
}

@misc{FDOH_BPHL_DENV_PCR_2023,
  title        = {Dengue virus (DENV) Typing, PCR},
 author       = {Florida Department of Health},
  institution  = {Florida Department of Health, Bureau of Public Health Laboratories},
  type         = {Test Menu},
  version      = {V1},
  date         = {2023-09},
  location     = {Jacksonville and Tampa, FL},
  url          = {https://www.floridahealth.gov/programs-and-services/public-health-laboratories/_documents/_test_menu/_documents/denv_pcr.pdf},
  urldate      = {2025-10-29},
  note         = {Test code 1681; DH 1847 requisition; store 2–8\degree C or $\leq$\,–20\degree C; ship on dry ice}
}

@misc{FDA_CLIA_2023,
  author       = {{U.S. Food and Drug Administration}},
  title        = {Clinical {L}aboratory {I}mprovement {A}mendments ({C}{L}{I}{A})},
  year         = {2023},
  howpublished = {\url{https://www.fda.gov/medical-devices/ivd-regulatory-assistance/clinical-laboratory-improvement-amendments-clia}},
  note         = {Accessed November 22, 2025}
}

@misc{CDC_Molecular_Tests_Dengue_2025,
  title       = {Molecular Tests for Dengue Virus},
  author      = {Centers for {D}isease {C}ontrol and {P}revention},
  date        = {2025-08-22},
  year        = {2025},
  url         = {https://www.cdc.gov/dengue/hcp/diagnosis-testing/molecular-tests-for-dengue-virus.html},
  urldate     = {2025-10-29},
  note        = {Guidance on NAAT/RT-PCR use for acute dengue diagnosis}
}

@misc{CDC_Clinical_Testing_Dengue_2025,
  title       = {Clinical Testing Guidance for Dengue},
  author      = {Centers for {D}isease {C}ontrol and {P}revention},
  date        = {2025-08-26},
  year        = {2025},
  url         = {https://www.cdc.gov/dengue/hcp/diagnosis-testing/index.html},
  urldate     = {2025-10-29},
  note        = {Guidance for clinicians on NAAT/RT-PCR, NS1, and IgM testing by illness day}
}

@article{chawla2014clinical,
  title={Clinical implications and treatment of dengue},
  author={Chawla, Pooja and Yadav, Amrita and Chawla, Viney},
  journal={Asian Pacific journal of tropical medicine},
  volume={7},
  number={3},
  pages={169--178},
  year={2014},
  publisher={Elsevier}
}

@article{rajapakse2012treatment,
  title={Treatment of dengue fever},
  author={Rajapakse, Senaka and Rodrigo, Chaturaka and Rajapakse, Anoja},
  journal={Infection and drug resistance},
  pages={103--112},
  year={2012},
  publisher={Taylor \& Francis}
}

@article{McKendrick,
author = {Kermack, William Ogilvy  and McKendrick, A. G.},
COMMauthor = {and Walker, Gilbert Thomas },
title = {A contribution to the mathematical theory of epidemics},
journal = {Proceedings of the Royal Society of London. Series A, Containing Papers of a Mathematical and Physical Character},
volume = {115},
number = {772},
pages = {700-721},
year = {1927},
doi = {10.1098/rspa.1927.0118},
COMMURL = {https://royalsocietypublishing.org/doi/abs/10.1098/rspa.1927.0118},
COMMeprint = {https://royalsocietypublishing.org/doi/pdf/10.1098/rspa.1927.0118}
}

@Inbook{Brauer2008,
author="Brauer, Fred",
editor="Brauer, Fred
and van den Driessche, Pauline
and Wu, Jianhong",
title="Compartmental Models in Epidemiology",
bookTitle="{Mathematical Epidemiology}",
year="2008",
publisher="Springer Berlin Heidelberg",
address="Berlin, Heidelberg",
pages="19--79",
isbn="978-3-540-78911-6",
doi="10.1007/978-3-540-78911-6_2",
COMMurl="https://doi.org/10.1007/978-3-540-78911-6_2"
}

@article{Filipe2003,
author = {J. A. N. Filipe and M. M. Maule},
title = {Analytical methods for predicting the behaviour of population models with general spatial interactions},
journal = {Mathematical Biosciences},
volume = {183},
pages = {15–35},
year = {2003},
doi = {10.1016/S0025-5564(02)00224-9},
}

@article{Cross2005,
author = {Paul C. Cross and James O. Lloyd-Smith and Philip L. F. Johnson and Wayne M. Getz},
title = {Duelling timescales of host movement and disease recovery determine invasion of disease in structured populations},
journal = {Ecology Letters},
volume = {183},
issue = {8},
pages = {587–595},
year = {2005},
doi = {10.1111/j.1461-0248.2005.00760.x},
}

@article{halstead2007dengue,
  title={Dengue},
  author={Halstead, Scott B},
  journal={The lancet},
  volume={370},
  number={9599},
  pages={1644--1652},
  year={2007},
  publisher={Elsevier}
}

@article{derouich2003model,
  title={A model of dengue fever},
  author={Derouich, M and Boutayeb, A and Twizell, EH},
  journal={Biomedical engineering online},
  volume={2},
  pages={1--10},
  year={2003},
  publisher={Springer}
}

@article{feng1997competitive,
  title={Competitive exclusion in a vector-host model for the dengue fever},
  author={Feng, Zhilan and Velasco-Hern{\'a}ndez, Jorge X},
  journal={Journal of mathematical biology},
  volume={35},
  pages={523--544},
  year={1997},
  publisher={Springer}
}

@article{side2013sir,
  title={A SIR model for spread of dengue fever disease (simulation for South Sulawesi, Indonesia and Selangor, Malaysia)},
  author={Side, Syafruddin and Noorani, Salmi Md},
  journal={World Journal of Modelling and Simulation},
  volume={9},
  number={2},
  pages={96--105},
  year={2013},
  publisher={Citeseer}
}

@article{nuraini2007mathematical,
  title={Mathematical model of dengue disease transmission with severe DHF compartment},
  author={Nuraini, N and Soewono, E and Sidarto, KA},
  journal={Bulletin of the Malaysian Mathematical Sciences Society},
  volume={30},
  number={2},
  year={2007},
  publisher={Springer Nature BV}
}

@article{usman2024analysis,
  title={Analysis of a fractional-order model for dengue transmission dynamics with quarantine and vaccination measures},
  author={Usman, Muhammad and Abbas, Mujahid and Khan, Safeer Hussain and Omame, Andrew},
  journal={Scientific Reports},
  volume={14},
  number={1},
  pages={11954},
  year={2024},
  publisher={Nature Publishing Group UK London}
}

@article{johnson2024investigating,
  title={Investigating the dose-dependency of the midgut escape barrier using a mechanistic model of within-mosquito dengue virus population dynamics},
  author={Johnson, Rebecca M and Stopard, Isaac J and Byrne, Helen M and Armstrong, Philip M and Brackney, Douglas E and Lambert, Ben},
  journal={PLoS Pathogens},
  volume={20},
  number={4},
  pages={e1011975},
  year={2024},
  publisher={Public Library of Science San Francisco, CA USA}
}

@article{xue2021transmission,
  title={Transmission dynamics of multi-strain dengue virus with cross-immunity},
  author={Xue, Ling and Zhang, Hongyu and Sun, Wei and Scoglio, Caterina},
  journal={Applied Mathematics and Computation},
  volume={392},
  pages={125742},
  year={2021},
  publisher={Elsevier}
}

@article{de2013stochastic,
  title={Stochastic dynamics of dengue epidemics},
  author={de Souza, David R and Tom{\'e}, T{\^a}nia and Pinho, Suani TR and Barreto, Florisneide R and de Oliveira, M{\'a}rio J},
  journal={Physical Review E—Statistical, Nonlinear, and Soft Matter Physics},
  volume={87},
  number={1},
  pages={012709},
  year={2013},
  publisher={APS}
}

@article{yu2014online,
  title={An online spatiotemporal prediction model for dengue fever epidemic in K aohsiung (T aiwan)},
  author={Yu, Hwa-Lung and Angulo, Jose M and Cheng, Ming-Hung and Wu, Jiaping and Christakos, George},
  journal={Biometrical Journal},
  volume={56},
  number={3},
  pages={428--440},
  year={2014},
  publisher={Wiley Online Library}
}

@article{lee2023comprehensive,
  title={Comprehensive analysis of multivariable models for predicting severe dengue prognosis: systematic review and meta-analysis},
  author={Lee, Hyelan and Hyun, Seungjae and Park, Sangshin},
  journal={Transactions of The Royal Society of Tropical Medicine and Hygiene},
  volume={117},
  number={3},
  pages={149--160},
  year={2023},
  publisher={Oxford University Press}
}

@article{pearl1920rate,
  title={On the rate of growth of the population of the United States since 1790 and its mathematical representation},
  author={Pearl, Raymond and Reed, Lowell J},
  journal={Proceedings of the national academy of sciences},
  volume={6},
  number={6},
  pages={275--288},
  year={1920}
}

@article{lega2016data,
  title={Data-driven outbreak forecasting with a simple nonlinear growth model},
  author={Lega, Joceline and Brown, Heidi E},
  journal={Epidemics},
  volume={17},
  pages={19--26},
  year={2016},
  publisher={Elsevier}
}

@article{lega2021parameter,
  title={Parameter estimation from ICC curves},
  author={Lega, Joceline},
  journal={Journal of biological dynamics},
  volume={15},
  number={1},
  pages={195--212},
  year={2021},
  publisher={Taylor \& Francis}
}

@article{sahneh2022epidemics,
  title={Epidemics from the Eye of the Pathogen},
  author={Sahneh, Faryad D and Fries, William and Watkins, Joseph C and Lega, Joceline},
  journal={SIAM Journal on Applied Mathematics},
  volume={82},
  number={6},
  pages={2036--2056},
  year={2022},
  publisher={SIAM}
}

@article{stephenson2021geographic,
  title={Geographic partitioning of dengue virus transmission risk in Florida},
  author={Stephenson, Caroline J and Coatsworth, Heather and Waits, Christy M and Nazario-Maldonado, Nicole M and Mathias, Derrick K and Dinglasan, Rhoel R and Lednicky, John A},
  journal={Viruses},
  volume={13},
  number={11},
  pages={2232},
  year={2021},
  publisher={MDPI}
}

@article{richards2012vector,
  title={Vector competence of Aedes aegypti and Aedes albopictus (Diptera: Culicidae) for dengue virus in the Florida Keys},
  author={Richards, Stephanie L and Anderson, Sheri L and Alto, Barry W},
  journal={Journal of medical entomology},
  volume={49},
  number={4},
  pages={942--946},
  year={2012},
  publisher={Oxford University Press Oxford, UK}
}

@article{holcomb2023evaluation,
  title={Evaluation of an open forecasting challenge to assess skill of West Nile virus neuroinvasive disease prediction},
  author={Holcomb, Karen M and Mathis, Sarabeth and Staples, J Erin and Fischer, Marc and Barker, Christopher M and Beard, Charles B and Nett, Randall J and Keyel, Alexander C and Marcantonio, Matteo and Childs, Marissa L and others},
  journal={Parasites \& Vectors},
  volume={16},
  number={1},
  pages={11},
  year={2023},
  publisher={Springer}
}

@misc{website-CDPH2025Dengue,
  title        = {Dengue Virus: What You Need to Know},
  author       = {{California Department of Public Health}},
  year         = {2025},
  howpublished = {\url{https://www.cdph.ca.gov/Programs/CID/DCDC/Pages/Dengue.aspx#:~:text=What\%20You\%20Need\%20to\%20Know,hours\%20and\%20requires\%20hospital\%20care.}},
  note         = {Accessed: 2025-07-31}
}

@misc{FloridaHealth2023,
    author       = {Florida Department of Health},
    title        = {Mosquito-Borne Diseases Surveillance},
    howpublished = {\url{https://www.floridahealth.gov/diseases-and-conditions/mosquito-borne-diseases/surveillance.html}},
    year         = {2023},
    note         = {[Last Modified Date: Aug 13, 2024 11:47:57 AM]},
publisher    = {Florida Department of Health}
}

@misc{nidss_cdc_tw,
  author       = {{Taiwan Centers for Disease Control}},
  title        = {Notifiable Infectious Disease Statistics System (NIDSS)},
  year         = {2024},
  url          = {https://nidss.cdc.gov.tw/en/nndss/disease?id=061},
  note         = {Accessed: 2024-09-04}
}

@article{thomas2019review,
  title={A review of Dengvaxia{\textregistered}: development to deployment},
  author={Thomas, Stephen J and Yoon, In-Kyu},
  journal={Human vaccines \& immunotherapeutics},
  volume={15},
  number={10},
  pages={2295--2314},
  year={2019},
  publisher={Taylor \& Francis}
}

@article{angelin2023qdenga,
  title={Qdenga{\textregistered}-A promising dengue fever vaccine; can it be recommended to non-immune travelers?},
  author={Angelin, Martin and Sj{\"o}lin, Jan and Kahn, Fredrik and Hedberg, Anna Ljunghill and Rosdahl, Anja and Skorup, Paul and Werner, Simon and Woxenius, Susanne and Askling, Helena H},
  journal={Travel Medicine and Infectious Disease},
  volume={54},
  pages={102598},
  year={2023},
  publisher={Elsevier}
}

@article{andraud2012dynamic,
  title={Dynamic epidemiological models for dengue transmission: a systematic review of structural approaches},
  author={Andraud, Mathieu and Hens, Niel and Marais, Christiaan and Beutels, Philippe},
  journal={PloS one},
  volume={7},
  number={11},
  pages={e49085},
  year={2012},
  publisher={Public Library of Science San Francisco, USA}
}

@article{yi2021seir,
  title={SEIR-SEI-EnKF: A new model for estimating and forecasting dengue outbreak dynamics},
  author={Yi, Chunlin and Cohnstaedt, Lee W and Scoglio, Caterina M},
  journal={IEEE Access},
  volume={9},
  pages={156758--156767},
  year={2021},
  publisher={IEEE}
}

@article{Nuraini2007,
  author  = {Nuraini and Edy Soewono and K.A. Sidarto},
  title   = {Mathematical Model of Dengue Disease Transmission with Severe DHF Compartment},
  journal = {Bulletin of the Malaysian Mathematical Sciences Society},
  volume  = {30},
  issue   = {2},
  pages   = {123--134},
  year    = {2007},
  doi     = {},}

@article{rey2014dengue,
  title={Dengue in Florida (USA)},
  author={Rey, Jorge R},
  journal={Insects},
  volume={5},
  number={4},
  pages={991--1000},
  year={2014},
  publisher={MDPI}
}

@article{stephenson2022imported,
  title={Imported dengue case numbers and local climatic patterns are associated with dengue virus transmission in Florida, USA},
  author={Stephenson, Caroline and Coker, Eric and Wisely, Samantha and Liang, Song and Dinglasan, Rhoel R and Lednicky, John A},
  journal={Insects},
  volume={13},
  number={2},
  pages={163},
  year={2022},
  publisher={MDPI}
}

@article{eisen2014temporal,
  title={Temporal correlations between mosquito-based dengue virus surveillance measures or indoor mosquito abundance and dengue case numbers in {M}erida {C}ity, {M}exico},
  author={Eisen, Lars and Garc{\'\i}a-Rej{\'o}n, Juli{\'a}n E and G{\'o}mez-Carro, Salvador and V{\'a}zquez, Mar{\'\i}a Del Rosario N{\'a}jera and Keefe, Thomas J and Beaty, Barry J and Loro{\~n}o-Pino, Mar{\'\i}a Alba},
  journal={Journal of medical entomology},
  volume={51},
  number={4},
  pages={885--890},
  year={2014},
  publisher={Oxford University Press Oxford, UK}
}

@article{liyanage2022assessing,
  title={Assessing the associations between Aedes larval indices and dengue risk in Kalutara district, {S}ri {L}anka: a hierarchical time series analysis from 2010 to 2019},
  author={Liyanage, Prasad and Tozan, Yesim and Tissera, Hasitha Aravinda and Overgaard, Hans J and Rockl{\"o}v, Joacim},
  journal={Parasites \& vectors},
  volume={15},
  number={1},
  pages={277},
  year={2022},
  publisher={Springer}
}

@article{da2017meteorological,
  title={Meteorological variables and mosquito monitoring are good predictors for infestation trends of Aedes aegypti, the vector of dengue, chikungunya and {Z}ika},
  author={da Cruz Ferreira, Danielle Andreza and Degener, Carolin Marlen and de Almeida Marques-Toledo, Cecilia and Bendati, Maria Mercedes and Fetzer, Liane Oliveira and Teixeira, Camila P and Eiras, {\'A}lvaro Eduardo},
  journal={Parasites \& vectors},
  volume={10},
  number={1},
  pages={78},
  year={2017},
  publisher={Springer}
}
\end{document}